\documentclass[preprint,12pt]{elsarticle}
\usepackage{amssymb}
\usepackage{svg}
\usepackage{amsmath}
\usepackage{makecell}
\usepackage{xcolor}
\usepackage{booktabs}
\usepackage{multirow}
\usepackage{caption}
\usepackage{subcaption}
\usepackage{amsmath}
\usepackage[table]{xcolor}
\usepackage{array}
\usepackage{longtable}
\usepackage{graphicx}
\usepackage[ruled,vlined]{algorithm2e}
\newcommand{\etal}{\textit{et al}. }
\usepackage[a4paper, left=2cm, right=2cm, top=2.5cm, bottom=2.5cm]{geometry}
\usepackage{microtype}
\usepackage{booktabs}
\usepackage{tabularx}
\usepackage[numbers]{natbib}
\definecolor{mypurple}{RGB}{114, 9, 183}
\journal{Nuclear Physics B}
\begin{document}
\begin{frontmatter}

\title{NeuroCLIP: Brain-Inspired Prompt Tuning for EEG-to-Image Multimodal Contrastive Learning}

\author{Jiyuan Wang, Xizhang Yao, Haipeng Lin, Ziyu Li, Yan Liu, Qile Liu, Gan Huang, Li Zhang, Zhen Liang, and Xia Wu}

% \author{Jiyuan Wang, Xizhang Yao, Li Zhang, Haipeng Lin, Qile Liu, Gan Huang\\ Ziyu Li, Zhen Liang, and Xia Wu}
%% Author name

% Author affiliation
% \affiliation{organization={The School of Biomedical Engineering},
%             addressline={Medical School}, 
%             city={Shenzhen University},
%             state={Shenzhen},
%             country={China}}
% \affiliation{organization={The Guangdong Provincial Key Laboratory of Biomedical Measurements and Ultrasound Imaging},
%             state={Shenzhen},
%             country={China}}
% \affiliation{organization={Shien-Ming Wu School of Intelligent Engineering,},
%             addressline={South China University of Technology}, 
%             state={Guangdong},
%             country={China}}
% \affiliation{organization={Institute for Super Robotics},
%             state={Guangdong},
%             country={China}}

% Abstract

\begin{abstract}
Recent advances in brain-inspired artificial intelligence have sought to align neural signals with visual semantics using multimodal models such as CLIP. However, existing methods often treat CLIP as a static feature extractor, overlooking its adaptability to neural representations and the inherent physiological-symbolic gap in EEG-image alignment. To address these challenges, we present \textbf{NeuroCLIP}, a prompt tuning framework tailored for EEG-to-image contrastive learning. Our approach introduces three core innovations: (1) We design a \textit{dual-stream visual embedding pipeline} that combines dynamic filtering and token-level fusion to generate \textit{instance-level adaptive prompts}, which guide the adjustment of patch embedding tokens based on image content, thereby enabling fine-grained modulation of visual representations under neural constraints; (2) We are the first to introduce \textit{visual prompt tokens} into EEG-image alignment, acting as global, modality-level prompts that work in conjunction with instance-level adjustments. These visual prompt tokens are inserted into the Transformer architecture to facilitate neural-aware adaptation and parameter optimization at a global level; (3) Inspired by neuroscientific principles of human visual encoding, we propose a \textit{refined contrastive loss} that better model the semantic ambiguity and cross-modal noise present in EEG signals. On the THINGS-EEG2 dataset, NeuroCLIP achieves a Top-1 accuracy of 63.2\% in zero-shot image retrieval, surpassing the previous best method by +12.3\%, and demonstrates strong generalization under inter-subject conditions (+4.6\% Top-1), highlighting the potential of physiology-aware prompt tuning for bridging brain signals and visual semantics.

\end{abstract}


\end{frontmatter}

% \footnotetext[1]{\hspace{1mm}Equal contributions.}

\section{Introduction}
\label{introduction}
%% Labels are used to cross-reference an item using \ref command.

Reading the brain and deciphering human consciousness has always been a fascinating topic. Visual decoding, which allows others to "see what you see", is a particularly interesting and mysterious technology in the frontier applications of brain-computer interfaces and neuro science. In recent decades, researchers have conducted extensive studies to investigate the mechanisms of the visual system of the brain\cite{li2023neural,xiao2023single,kanwisher1997fusiform,treue1999feature}. However, the semantic understanding of natural images by the human brain’ s visual system remains a central mystery in cognitive science, with decoding visual information from electroencephalogram (EEG) signals offering a critical technological avenue to unravel this process\cite{liu2024visual}. EEG’s millisecond-level temporal resolution, low cost, and portability position it as an ideal tool for real-time brain-computer interfaces (BCIs) and neural decoding\cite{lalor2015decoding}. However, its low signal-to-noise ratio (SNR)\cite{luo2020eeg}, limited spatial resolution\cite{wendel2009eeg}, and the nonlinear mapping gap between visual stimuli and neural responses\cite{pan2024reconstructing} pose substantial challenges for extracting stable semantic features, particularly in zero-shot object recognition, where models must generalize to unseen categories, exposing the limitations of traditional supervised learning on scarce labeled data\cite{song2024decoding}.

Beyond its scientific importance in dissecting brain mechanisms, EEG-based visual decoding serves as a cornerstone for advancing non-invasive BCI applications in industrial and daily contexts\cite{varbu2022past}. In healthcare, it enables non-invasive visual function assessment via EEG signal analysis, providing objective biomarkers for diagnosing and rehabilitating neurological disorders like autism and stroke. For example, decoding visual-evoked EEG signals in Alzheimer’s patients may detect early degeneration in their visual-semantic networks. In intelligent interaction, real-time decoding empowers "thought-controlled" devices—such as wheelchair navigation or smart home systems—by recognizing neural signatures of object fixation, overcoming limb-dependent input limitations.

In the deep learning era, EEG-based visual decoding has emerged as a unique branch of multimodal learning. This line of research seeks to decode human visual perception—such as object categories or scene semantics, by analyzing non-invasive EEG signals evoked by visual stimuli, enabling downstream tasks such as classification, retrieval, and even reconstruction through cross-modal learning techniques like contrastive or generative modeling~\cite{wei2024mb2c,guo2024neuro,chen2024visual,chen2024muse,bai2023dreamdiffusion,lan2023seeing}.
However, despite promising progress, two fundamental challenges remain unresolved:
\textbf{(1) The Human–Computer Perception Gap in EEG Decoding.}
As first formally defined in UBP~\cite{wu2025bridging}, this gap arises from hardware limitations (e.g., low spatial resolution, noisy signals) and distortions introduced by preprocessing. Consequently, frequency and temporal features extracted from EEG are often misaligned with actual perceptual content, complicating semantic interpretation.
\textbf{(2) The Cross-Modal Misalignment Between EEG and Images.}
Unlike text-image pairs where symbolic structures allow direct alignment, EEG signals encode visual information implicitly as latent neural activity distributed across time and frequency. The lack of explicit, human-annotated intermediate representations leads to a profound "physiological-symbolic" modality gap that poses significant challenges for learning effective EEG-image correspondences.
While a few recent studies have explored the use of multimodal foundation models, such as the CLIP family~\cite{radford2021learning}, to bridge the gap between EEG signals and visual semantics, these approaches often underutilize the full capacity of such models. In most cases, pretrained vision-language models are treated as static feature extractors, with little to no task-specific adaptation or modality-aware tuning. This practice overlooks the fundamental mismatch between the neural representations captured by EEG signals and the semantic structures encoded in image-text pairs. As a result, existing methods fall short in fully leveraging the representational power of pretrained multimodal encoders in the context of EEG-image alignment. These limitations highlight the need for more flexible fine-tuning of pretrained multimodal models to better accommodate new modalities and tasks.

Among various fine-tuning strategies for adapting pretrained multimodal models to new tasks and modalities, prompt tuning has emerged as a lightweight yet effective solution. Instead of updating the full model, prompt tuning introduces a small set of learnable tokens that are appended to the input sequence to guide the model’s behavior. Notable examples include CoOp~\cite{zhou2022learning}, which learns continuous prompts in the text encoder of CLIP for improved visual classification, and VPT (Visual Prompt Tuning)~\cite{jia2022vpt}, which extends this idea to the visual encoder by prepending learnable prompts to image patch tokens. These methods have demonstrated impressive performance with minimal parameter tuning, making prompt learning a compelling tool for efficient task adaptation.
However, existing prompt tuning techniques typically adopt a static and unidirectional design, where prompt tokens are simply appended to fixed input embeddings—such as patch embeddings for images or token embeddings for text—without updating the underlying features during training. The adaptation occurs only through the prompt tokens, while the input representations remain unchanged. Conceptually, this overlooks the interactive nature of guided learning. Much like a teacher guiding a student, effective prompting should be a two-way process: while the prompts shape the model’s output, the input representations themselves should also evolve in response to the guidance. This limitation calls for a more dynamic and bidirectional formulation of prompt tuning, particularly in the context of EEG-image alignment, where the modality gap is inherently large.

\begin{figure}[htbp]
    \centering
    \includegraphics[width=\textwidth]{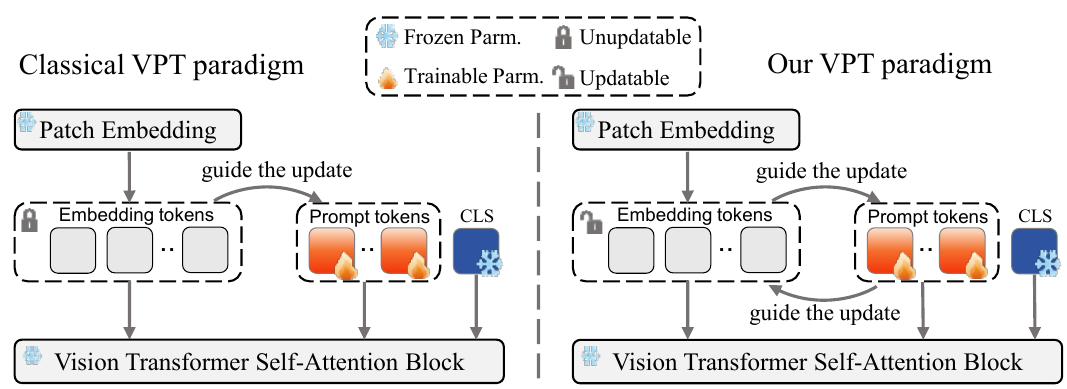}
    \caption{Comparison between the Classical and Our Proposed Visual Prompt Tuning Paradigms}
    \label{fig:methodinshort}
\end{figure}

Building upon the aforementioned limitations, we propose NeuroCLIP, a novel and flexible EEG-image alignment framework that fine-tunes CLIP-based vision models to better accommodate neural data. Our framework features a more adaptable alignment architecture, adopts a more efficient prompt tuning strategy, and introduces improvements to the original CLIP cross-modal alignment loss. This design enables the pretrained multimodal encoder, which was originally optimized for image-text pairing, to be effectively adapted to the EEG modality, thereby enhancing its ability to model and align brain signals with visual semantics. Our key contributions are as follows:

To address the core challenges discussed earlier, we propose NeuroCLIP, a novel framework that leverages adaptive prompt learning to enhance cross-modal alignment between EEG and visual stimuli. Unlike prior approaches that mechanically incorporate pre-trained multimodal models (such as CLIP-VIT) into the framework without task-specific adjustments, our method introduces a dynamic visual prompt tuning technique to fine-tune the CLIP-VIT model, ensuring better adaptation to EEG data and significantly improving performance on downstream tasks. Figure~\ref{fig:methodinshort} demonstrates the differences between the classical visual prompt tuning paradigm and the more flexible visual prompt tuning paradigm proposed by us.
Our key contributions are as follows:

%To address the core technical challenges outlined above, we propose NeuroCLIP, a novel framework leveraging adaptive prompt learning. Our approach enables the visual encoder to dynamically learn feature representations of visual inputs guided by EEG signals, generating self-learned visual prompts that are integrated into the fine-tuning of the CLIP-VIT model. This endows the pre-trained CLIP-VIT—originally optimized on large-scale image-text pairs—with an enhanced capability to bridge the modality gap: it now captures the deep semantic correlations between neural signals (EEG) and visual content more effectively than standard CLIP models, improving cross-modal understanding of brainwave data. In summary, our key contributions are as follows:

(1)We propose NeuroCLIP, a novel and flexible framework for EEG-image cross-modal alignment. The framework improves upon conventional designs in three key aspects: the efficiency of modality-specific feature encoding, the flexibility of prompt tuning strategies, and the design of a more effective alignment loss. These components are jointly optimized to achieve better multimodal alignment performance.

(2)We introduce a more efficient and flexible visual prompt tuning strategy. To the best of our knowledge, this is the first work to apply visual prompt tuning to fine-tune CLIP-VIT models in the context of EEG-image alignment. Unlike the standard Visual Prompt Tuning (VPT) paradigm, our method introduces a two-level prompting strategy, consisting of Instance-level Visual Prompt Tuning(IVPT) and Shared-level Visual Prompt Tuning(SVPT). This design enhances bidirectional interaction and co-adaptation between prompt parameters and the target visual features, making the tuning process more effective. Figure~\ref{fig:methodinshort} demonstrates the differences between the standard visual prompt tuning paradigm and the more flexible visual prompt tuning paradigm proposed by us.

(3)We improve the original CLIP alignment objective by designing a more principled cross-modal loss tailored to EEG-visual representation learning. Inspired by the visual encoding mechanisms of the human brain, our loss reformulates the traditional contrastive objective by relaxing the hard-label constraint, enabling more stable and semantically consistent correlations between EEG signals and image features. This adaptation enhances the model’s robustness and generalization across cross-modal tasks.

(4)Our proposed model, NeuroCLIP, achieves state-of-the-art performance on the 200-way zero-shot EEG-image cross-modal retrieval task of the THINGS-EEG2 benchmark, under both intra-subject and inter-subject evaluation settings.

%methodinshort

\section{Related work}
%% Use \subsection commands to start a subsection.
\textbf{EEG based visual decoding}

Researchers have been exploring the visual coding mechanisms of the brain since early days, and have endeavored to decipher the visual information embedded within the acquired electroencephalogram (EEG) signals.In the early days of cognitive neuroscience, researchers utilized event-related potential (ERP) and steady-state visual evoked potential (SSVEP) analyses to preliminarily uncover the neurophysiological encoding rules of visual stimuli from a mechanistic perspective\cite{hillyard1998event,farwell1988talking,carlson2011high,wang2012combining}. There were only a few works that conducted some very simple visual classification tasks (most of which were only binary classification tasks) based on the theoretical foundations of these early visual decoding studies\cite{bigdely2008brain,kapoor2008combining,stewart2014single}. These studies generally did not employ machine learning or deep learning models. Moreover, the classification tasks were relatively simple, involving a small number of categories and limited types of visual stimuli (with a lack of natural visual scene stimuli). Later, with the development of artificial intelligence algorithms, some methods began to attempt to construct machine learning or deep learning classifiers to complete more complex classification tasks. Among them, the most representative research work is from Spampinato \etal \cite{spampinato2017deep}. They translate the acquired capabilities to machines by training a Convolutional Neural Network (CNN)-based regressor to map images onto the learned manifold, thereby enabling machines to leverage human brain-derived features for automated visual classification(40 object classes). These works have laid a valuable theoretical foundation and accumulated practical experience for EEG-based visual decoding tasks, but their limitations cannot be ignored— the single type of visual stimuli in the dataset and the insufficient generalization ability of the models make it difficult to adapt to different visual scene stimuli. This, in turn, results in the models' inability to capture the visual encoding mechanisms of EEG signals in complex real-world scenarios.

% \textbf{CLIP}

% The original CLIP model\cite{radford2021learning} revolutionized vision-language pre-training through contrastive learning with large-scale image-text pairs. However, its limitations in data efficiency and class generalization spurred subsequent advancements. PyramidCLIP \cite{gao2022pyramidclip} improved alignment via hierarchical feature structures (e.g., ROI features and tags), achieving significant accuracy gains on tasks like ImageNet\cite{deng2009imagenet}. CoOp\cite{zhou2022learning} and CoCoOp\cite{zhou2022conditional} optimized learnable context tokens for prompt adaptation, with CoCoOp enhancing generalization to unseen classes. SoftCLIP\cite{gao2024softclip} introduced softened supervision (e.g., ROI-to-tag similarity) to mitigate noisy data, while RankCLIP\cite{zhang2024rankclip} integrated ranking consistency loss to capture relative semantic similarities, improving robustness to distribution shifts. These works collectively address CLIP’s rigidity, emphasizing task-aware designs (e.g., dynamic prompts, soft alignment, and ranking-based losses) for scalable real-world applications. 

\textbf{EEG-Image retrival}

Visual decoding from EEG has advanced through cross-modal learning. Song \etal introduced NICE, using CLIP and spatial-temporal convolutions with attention (SA/GA) to exploit electrode correlations, improving retrieval accuracy\cite{song2023decoding}. Chen \etal further proposed MUSE, enhancing contrastive learning with similarity-keeping losses to refine latent-space consistency\cite{chen2024mind}. Li \etal employed an improved EEG encoder, the Adaptive Thinking Mapper (ATM), achieving a top-1 accuracy of 28.64\% in retrieval tasks\cite{li2024visual}. Chen \etal proposed VE-SDN to decouple semantic features from visual images and EEG signals for cross-modal alignment \cite{chen2024visual}. Rajabi \etal developed human-aligned image encoders (e.g., Dreamsim), fine-tuned on human perception data to match rapid neural responses (100ms), boosting retrieval accuracy\cite{rajabi2025human}. Wu \etal addressed modality gaps with UBP, a blur prior adaptive to neural uncertainty, which achieved the best performance at the time \cite{wu2025bridging}.

\section{Methods}

\begin{figure}[htbp]
    \centering
    \includegraphics[width=1\textwidth]{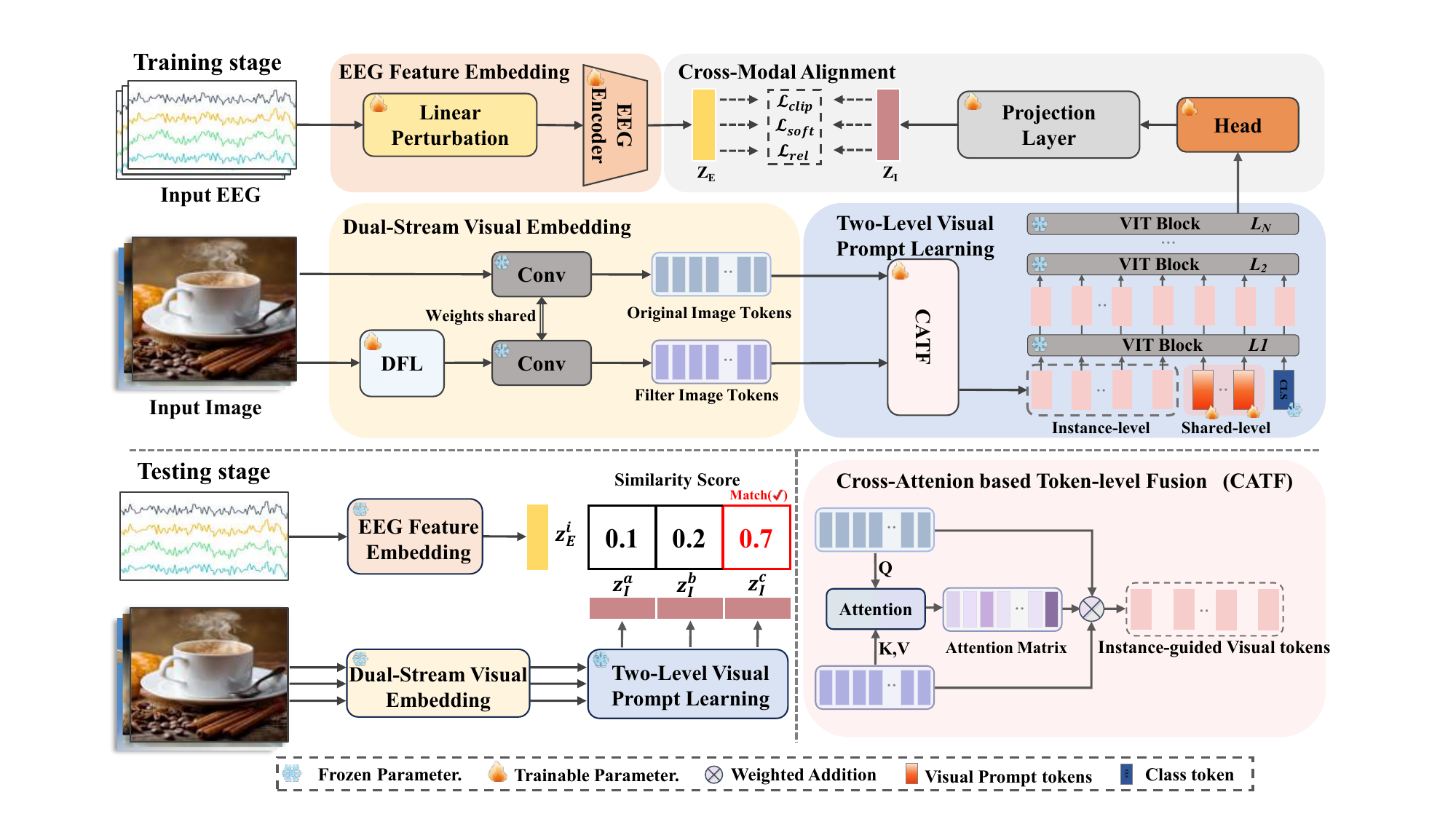}
    \caption{The NeuroCLIP framework. EEG signals are perturbed and encoded; Images are processed through a Dual-Stream Visual Embedding with a Dynamic Filter Layer (DFL). Instance-specific cues are injected by Cross-Attention Token-level Fusion (CATF), and Two-Level Visual Prompt Learning introduces both instance-level and shared-level prompts into the frozen CLIP-VIT. EEG–Image embeddings are then projected and aligned for cross-modal retrieval.}
    \label{fig:overallstructure}
\end{figure}

We introduce NeuroCLIP, a novel fine-tuning framework for CLIP-VIT tailored to EEG–image alignment tasks. The proposed framework comprises the following key components: (1) EEG Feature Embedding Module: This module is designed to extract informative and compact representations from raw EEG signals. (2) Dual-Stream Visual Embedding Module: Unlike the single-stream patch embedding architecture adopted in standard CLIP-VIT, our framework employs a dual-stream patch embedding design. Specifically, we extract embedding representations from both the original image and its adaptively filtered counterparts, which are generated via a content-adaptive dynamic filtering process, providing complementary visual features for subsequent fusion. (3) Two Level Visual Prompt Learning Module: In this module, the embedding of the dynamically filtered images serves as instance-level prompts. A cross-attention-based token-level fusion mechanism is introduced to dynamically adjust the patch token embeddings of the original image based on these instance-level prompts. Additionally, shared-level prompt tokens across samples are incorporated, which interact bidirectionally with the dynamically adjusted patch tokens during training. Within the CLIP-based alignment framework, both instance-level and shared-level prompt tokens are optimized to enhance the quality of image representations and improve their alignment with EEG features. (4) Cross-Modal Alignment Loss: Inspired by neuroscientific principles of visual encoding, this loss function softens the conventional CLIP contrastive objective by reducing reliance on hard semantic label supervision. Instead, it emphasizes the alignment between similar visual stimuli and their corresponding EEG representations, thereby enhancing the model’s ability to capture fine-grained cross-modal correlations. The overall architecture is illustrated in Figure~\ref{fig:overallstructure}.

\subsection{EEG Feature Embedding Module}
To obtain encoded EEG features for multimodal alignment with image representations, we first apply a \textbf{learnable linear perturbation module} for signal augmentation. Given raw EEG input $\mathbf{E} \in \mathbb{R}^{C \times T}$, the perturbation enhances signal expressiveness by applying element-wise affine transformations:
\begin{equation}
\hat{\mathbf{E}} = \mathbf{E} \odot \mathbf{W} + \mathbf{B}
\end{equation}
where $\mathbf{W}, \mathbf{B} \in \mathbb{R}^{C \times T}$ are trainable weights and biases initialized as identity. This lightweight operation allows the model to adaptively emphasize or suppress specific spatiotemporal regions of the EEG signal.

EEG signals are known to be non-stationary and low in signal-to-noise ratio, with informative neural patterns such as event-related potentials (ERPs) often buried within background activity. The learnable perturbation helps highlight such task-relevant components, improving the expressiveness of the extracted features. Additionally, this operation serves as a data-driven alternative to conventional EEG augmentation, enabling the network to learn transformation patterns that support better generalization and more robust cross-modal alignment.

After perturbation, we encode the modified EEG signals $\hat{\mathbf{E}}$ using \textbf{several alternative lightweight EEG encoders}, each integrated independently to assess their effectiveness within our framework. These encoder choices follow classical designs that have been frequently adopted in prior EEG-based visual decoding studies. This setting facilitates fair comparisons and controlled benchmarking. Detailed encoder configurations and performance comparisons are provided in the experimental section.

\subsection{Dual-Stream Visual Embedding Module}

% In contrast to conventional CLIP-based pipelines that adopt a single-branch vision encoder, our model introduces a dual-branch image encoding strategy designed to enhance the adaptability of visual representations in multimodal alignment tasks. Specifically, one branch directly encodes the raw image patches, while the second branch generates a dynamically filtered version of the same image, conditioned on visual content and optimized through training to align better with EEG features.

% The filtered and original representations are then fused through a token-level cross-attention mechanism, which adaptively adjusts each patch token before passing them into the VIT backbone. This distribution-aware fusion enables the model to learn more expressive, cross-modal-aligned visual tokens by treating the dynamically filtered branch as a content-aware adaptor or prompt that guides and refines the original features.

% This dual-branch architecture thus serves a dual purpose: while one path retains the fidelity of raw visual semantics, the other learns to reshape those semantics under multimodal constraints—leading to more discriminative and alignment-aware visual embeddings for downstream tasks such as EEG-to-image retrieval and classification.

In contrast to conventional CLIP-based pipelines that adopt a single-stream visual embedding method, our model introduces a dual-stream visual embedding strategy. While one stream preserves the original visual semantics, the other generates alignment-aware features by incorporating feedback from the cross-modal training process. These features, containing useful guidance signals, are used in the next module to adaptively modulate the original visual embeddings. To generate such features, we draw inspiration from prior work~\cite{jia2016dynamic} and design a mechanism that adaptively constructs alignment-aware representations based on visual content. Specificly, we introduce a dynamic filtering mechanism that generates and applies content-adaptive filters based on the input image itself. Given an input image $\mathbf{I} \in \mathbb{R}^{3 \times H \times W}$, a lightweight CNN followed by a multi-layer perceptron is used to produce a compact filter representation tailored to the visual content.

Let $F_h$ and $F_w$ denote the height and width of the filter kernel (typically $5 \times 5$). For RGB input, the total number of filter parameters is $3 \times F_h \times F_w$. The filter generator outputs:
\begin{equation}
\mathbf{f} = \phi(\mathbf{I}) \in \mathbb{R}^{3 \cdot F_h \cdot F_w}
\end{equation}
The resulting filter $\mathbf{f}$ is then applied to the input image through a dynamic filtering layer, which performs local patch unfolding and channel-wise convolution, producing a filtered image $\mathbf{I}_{\text{filt}}$:
\begin{equation}
\mathbf{I}_{\text{filt}} = \text{DynamicFilterLayer}(\mathbf{I}, \mathbf{f})
\end{equation}
This filtered output retains spatially adaptive characteristics reflective of the input, thereby enriching the image representation with instance-specific structural information. 

Next, we perform parallel feature embedding on the original image $\mathbf{I}$ and its dynamically filtered counterpart $\mathbf{I}_{\text{filt}}$. Specifically, both $\mathbf{I}$ and $\mathbf{I}_{\text{filt}}$ are passed through a shared patch embedding layer, which corresponds to the first convolutional stem of a pre-trained CLIP-VIT encoder. The parameters of this layer are frozen during training to preserve the original visual feature priors and ensure consistency across streams. This yields two sets of patch-level embeddings:
\begin{align}
\mathbf{X}_{\text{orig}} &= \text{PatchEmbed}(\mathbf{I}) \in \mathbb{R}^{N \times d} \\
\mathbf{X}_{\text{filt}} &= \text{PatchEmbed}(\mathbf{I}_{\text{filt}}) \in \mathbb{R}^{N \times d}
\end{align}
where $N$ is the number of patch tokens and $d$ is the embedding dimension.

%detaildynamicfilter
% \begin{table}[h]
% \centering
% \caption{Dynamic Filter Layer Configuration}
% \label{tab:dynamic_filter_layer}
% \begin{tabular}{l|l}
% \toprule
% \textbf{Component} & \textbf{Details} \\
% \midrule
% Input Image & $\mathbb{R}^{B \times 3 \times H \times W}$ \\
% Filter Shape & $\mathbb{R}^{B \times (3 \cdot F_h \cdot F_w)}$ \\
% Expansion & Local patch unfolding (per channel) \\
% Filtering & Element-wise multiply + sum over spatial patch \\
% Output & Filtered image $\in \mathbb{R}^{B \times 3 \times H \times W}$ \\
% \bottomrule
% \end{tabular}
% \end{table}

\subsection{Image Prompt Learning Module}
\subsubsection{Instance-level Visual Prompt Tunning(IVPT)}
\label{sec:fusion}
We further leverage $\mathbf{X}_{\text{filt}}$ as an instance-level visual prompt to provide sample-specific guidance for refining $\mathbf{X}_{\text{orig}}$. Rather than naively aggregating the two views, we aim to apply this prompt adaptively while preserving the spatial specificity of the original image tokens. To this end, we propose a \textbf{Cross-Attention Based Token-Level Fusion(CATF)} strategy that performs token-wise cross-attention from $\mathbf{X}_{\text{orig}}$ to $\mathbf{X}_{\text{filt}}$. Specifically, the details of CATF are as follows. In the cross-attention operation, we use $\mathbf{X}_{\text{orig}}$ as the queries and $\mathbf{X}_{\text{filt}}$ as the keys and values:
\begin{align}
\mathbf{Q} &= \mathbf{X}_{\text{orig}} \mathbf{W}^Q,\quad 
\mathbf{K} = \mathbf{X}_{\text{filt}} \mathbf{W}^K,\quad 
\mathbf{V} = \mathbf{X}_{\text{filt}} \mathbf{W}^V \\
\mathbf{Z} &= \text{Softmax}\left( \frac{\mathbf{Q} \mathbf{K}^\top}{\sqrt{d}} \right) \mathbf{V}
\end{align}
A fusion gate then determines how much each token should be influenced by the filtered branch:
\begin{equation}
\boldsymbol{\alpha} = \sigma(\text{FFN}(\mathbf{Z})) \in \mathbb{R}^{N \times 1}
\end{equation}
The final fused token representation becomes:
\begin{equation}
\mathbf{X}_{\text{fused}} = \boldsymbol{\alpha} \odot \mathbf{X}_{\text{filt}} + (1 - \boldsymbol{\alpha}) \odot \mathbf{X}_{\text{orig}}
\end{equation}

CATF enables the instance-level prompt tokens to effectively guide the modulation of the original visual embeddings, facilitating adaptive and fine-grained integration of instance-specific visual content. For completeness, we also compare CATF  with a simpler baseline fusion strategy, \textbf{Bilinear Adapter Fusion}, which performs a learnable convex combination of
the original and filtered images:
\begin{equation}
\mathbf{I}_{\text{mix}} = \lambda \cdot \mathbf{I}_{\text{filt}} + (1 - \lambda) \cdot \mathbf{I}, \quad \lambda \in (0, 1)
\label{eq:filtered_mix}
\end{equation}
However, as shown in later experiments, our proposed token-level method consistently achieves superior alignment and retrieval performance.

\subsubsection{Shared-level Visual Prompt Tuning(SVPT)}

Building on our Dual-Stream Visual Embedding Module and Instance-level Visual Prompt Tuning strategy, which yields high-quality patch tokens $\mathbf{X}_{\text{fused}}$, we further enhance the model’s adaptability by incorporating \textbf{Shared-level Visual Prompt Tuning (SVPT)}. The instance-level prompts are generated in a one-to-one manner for each individual image sample, allowing the patch embeddings to be dynamically modulated based on input-specific context. In contrast, the shared-level prompt tokens are globally learned and shared across all samples, serving as a unified and task-adaptive semantic prior. These shared prompts are injected into the visual transformer alongside the instance-specific patch tokens, enabling bi-directional interaction through the self-attention mechanism. Unlike conventional VPT approaches (e.g.,~\cite{jia2022visual}), our SVPT benefits from the joint optimization of shared prompts and dynamically modulated visual tokens, leading to a more flexible and semantically aligned prompting mechanism.

Specifically, we introduce a set of learnable prompt tokens $\mathbf{P} \in \mathbb{R}^{N_p \times d}$, where $N_p$ is the number of prompt tokens. These are inserted into the input sequence of the CLIP visual transformer between the [CLS] token and $\mathbf{X}_{\text{fused}}$, forming:
\begin{equation}
\mathbf{X}_{\text{VIT}} = [\mathbf{x}_{\text{cls}}; \mathbf{P}; \mathbf{X}_{\text{fused}}] \in \mathbb{R}^{(1 + N_p + N) \times d}
\end{equation}
Corresponding positional embeddings are expanded and added accordingly to maintain token semantics.

$\mathbf{X}_{\text{VIT}}$ is then fed into the pretrained CLIP-VIT blocks $f_V$ , whose parameters are kept frozen throughout training. During this process, the instance-level prompt-modulated image tokens and the dynamically updated shared-level prompt tokens interact through the self-attention mechanism within the VIT architecture. These interactions enable dynamic control over the model's final representation while preserving the visual prior encoded in the pretrained backbone. The output token corresponding to the [CLS] position is denoted as $\mathbf{z}_I^{\text{VIT}} = f_V(\mathbf{X}_{\text{VIT}})_{\text{cls}}$, representing the global visual representation under prompt guidance. To align with the EEG feature space, this representation is further projected through a trainable linear projecton layer (MLP), yielding the final image embedding:
\begin{equation}
\mathbf{z}_I = \text{MLP}(\mathbf{z}_I^{\text{VIT}}) \in \mathbb{R}^{d}
\end{equation}
This projection ensures that the image features are mapped into a shared latent space consistent with the EEG embeddings, facilitating effective cross-modal alignment.

Our motivation stems from the observation that the filtered-and-fused patch tokens already encode context-aware, EEG-relevant information. By inserting prompt tokens after the [CLS] token, we encourage the model to treat them not as standalone cues, but as \textit{collaborative adaptors} that work in tandem with already informative patch tokens.
% This setup allows:
% \begin{itemize}
%     \item \textbf{Prompt–Token Synergy}: fused tokens dynamically influence and are influenced by prompt tokens, enabling bidirectional adaptation.
%     \item \textbf{Task-Aware Adaptation}: unlike static prompt tuning, our model continuously adjusts both prompts and patch tokens to better fit the multimodal alignment task.
%     \item \textbf{Position-Aware Modulation}: placing prompts between [CLS] and image tokens allows them to capture upstream global attention while guiding downstream semantic aggregation.
% \end{itemize}
This hybrid design provides stronger generalization than traditional prompt tuning, especially in scenarios involving non-textual modalities such as EEG.

\subsection{Cross-Modal Alignment Loss}

Traditional contrastive learning objectives, such as the one used in CLIP, are based on the InfoNCE loss~\cite{oord2018representation}. These approaches assume that for each input query, there exists exactly one positive target and all other examples in the batch are negatives. However, in the context of EEG-based image alignment, especially under rapid serial visual presentation (RSVP), this hard assumption may not hold.

According to the conclusions of previous studies~\cite{rajabi2025human}, during RSVP, the human brain may not fully encode high-level semantic categories. Instead, EEG signals are more likely to reflect lower- or mid-level visual attributes such as object shape, color, or layout. As a result, treating non-matching image–EEG pairs as entirely unrelated (i.e., negatives) ignores their potential perceptual similarity.

To better model this ambiguity, we introduce a \textbf{softened contrastive objective} that incorporates intra-modal relational structure as soft targets.

\paragraph{Original Contrastive Loss (InfoNCE Formulation)}

Let $\mathbf{E}_i, \mathbf{I}_j \in \mathbb{R}^d$ denote the normalized embeddings of the $i$-th EEG sample and the $j$-th image sample. Their cosine similarity is computed as:
\begin{equation}
\text{sim}(\mathbf{E}_i, \mathbf{I}_j) = \frac{\mathbf{E}_i^\top \mathbf{I}_j}{\|\mathbf{E}_i\| \cdot \|\mathbf{I}_j\|}
\end{equation}

Given a batch of $B$ samples, the InfoNCE loss used in CLIP is defined as:
\begin{equation}
\mathcal{L}_{\text{clip}} = \frac{1}{2B} \sum_{i=1}^{B} \left[ 
- \log \frac{\exp(\text{sim}(\mathbf{E}_i, \mathbf{I}_i)/\tau)}{
\sum_{j=1}^{B} \exp(\text{sim}(\mathbf{E}_i, \mathbf{I}_j)/\tau)
}
- \log \frac{\exp(\text{sim}(\mathbf{I}_i, \mathbf{E}_i)/\tau)}{
\sum_{j=1}^{B} \exp(\text{sim}(\mathbf{I}_i, \mathbf{E}_j)/\tau)
}
\right]
\end{equation}
where $\tau$ is a learnable temperature parameter.

\paragraph{Soft Target Formulation}

To move beyond binary supervision, we compute intra-modal similarity distributions using softmax:
\begin{equation}
\mathbf{P}_{\text{EE}} = \text{Softmax}\left( \frac{\mathbf{E} \mathbf{E}^\top}{\tau} \right), \quad 
\mathbf{P}_{\text{II}} = \text{Softmax}\left( \frac{\mathbf{I} \mathbf{I}^\top}{\tau} \right)
\end{equation}

We define soft targets as a convex interpolation between the identity matrix and intra-modal similarity:
\begin{equation}
\mathbf{T}_{\text{E}} = (1 - \beta) \cdot \mathbf{I} + \beta \cdot \mathbf{P}_{\text{EE}}, \quad
\mathbf{T}_{\text{I}} = (1 - \beta) \cdot \mathbf{I} + \beta \cdot \mathbf{P}_{\text{II}}
\end{equation}

The soft alignment loss is formulated using symmetric KL divergence:
\begin{align}
\mathcal{L}_{\text{soft}} &= \frac{1}{2} \left[ D_{\text{KL}}(\mathbf{T}_{\text{E}} \| \mathbf{P}_{\text{EI}}) + D_{\text{KL}}(\mathbf{P}_{\text{EI}} \| \mathbf{T}_{\text{E}}) \right] \nonumber \\
&\quad + \frac{1}{2} \left[ D_{\text{KL}}(\mathbf{T}_{\text{I}} \| \mathbf{P}_{\text{IE}}) + D_{\text{KL}}(\mathbf{P}_{\text{IE}} \| \mathbf{T}_{\text{I}}) \right]
\end{align}
where $\mathbf{P}_{\text{EI}}$ and $\mathbf{P}_{\text{IE}}$ are cross-modal similarity distributions.

\paragraph{Relation-Aware Regularization}

To encourage separation of similar negatives, we disentangle the probability distributions by removing diagonal elements (positives), renormalize, and compute relation-enhancement loss:
\begin{equation}
\mathcal{L}_{\text{rel}} = \frac{1}{2} \left[ 
D_{\text{KL}}(\text{neg}(\mathbf{P}_{\text{EE}}) \| \text{neg}(\mathbf{P}_{\text{EI}})) +
D_{\text{KL}}(\text{neg}(\mathbf{P}_{\text{II}}) \| \text{neg}(\mathbf{P}_{\text{IE}})) 
\right]
\end{equation}

\paragraph{Final Objective}

The total loss is a weighted combination:
\begin{equation}
\mathcal{L}_{\text{total}} = \mu \cdot \mathcal{L}_{\text{clip}} + \alpha \cdot \mathcal{L}_{\text{soft}} + \lambda \cdot \mathcal{L}_{\text{rel}}
\end{equation}

\vspace{0.5em}
 This design is inspired by recent advances in soft contrastive learning from the vision–language domain~\cite{gao2024softclip}. Our work is the first to extend this technique to EEG–Image multimodal alignment, where semantic uncertainty and perceptual correlation between “negatives” are especially prominent. Algorithm~\ref{alg:neuroclip}  introduces the overall algorithmic process of NeuroCLIP.

%algorithm table
\begin{algorithm}[t]
\caption{Training Processing of NeuroCLIP Framework}
\label{alg:neuroclip}
\KwIn{Paired EEG-image dataset $\mathcal{D} = \{(\mathbf{E}_i, \mathbf{I}_i)\}_{i=1}^N$, where $\mathbf{E}_i \in \mathbb{R}^{C \times T}$, $\mathbf{I}_i \in \mathbb{R}^{3 \times H \times W}$}
\KwOut{Trained EEG encoder $f_E$, learnable perturbation $(\mathbf{W}, \mathbf{B})$, dynamic filter generator $\phi$, cross-attention based token-level fusion layer $\text{CATF}$, shared-level prompt tokens $\mathbf{P}$, and projection head MLP}

\ForEach{mini-batch $(\mathbf{E}, \mathbf{I}) \subset \mathcal{D}$}{
    \tcp{\textbf{EEG Feature Embedding Module (Trainable: $f_E$, $\mathbf{W}, \mathbf{B}$)}}
    $\hat{\mathbf{E}} \leftarrow \mathbf{E} \odot \mathbf{W} + \mathbf{B}$ \tcp*[f]{Learnable perturbation on EEG} \\
    $\mathbf{z}_E \leftarrow f_E(\hat{\mathbf{E}})$ \tcp*[f]{EEG encoder projects into shared space} \\

    \tcp{\textbf{Dual-Stream Visual Embedding Module (Trainable: $\phi$)}}
    $\mathbf{f} \leftarrow \phi(\mathbf{I})$ \tcp*[f]{Generate dynamic filters} \\
    $\mathbf{I}_{\text{filt}} \leftarrow \text{DynamicFilterLayer}(\mathbf{I}, \mathbf{f})$ \\
    $\mathbf{X}_{\text{orig}} \leftarrow \text{PatchEmbed}(\mathbf{I})$, $\mathbf{X}_{\text{filt}} \leftarrow \text{PatchEmbed}(\mathbf{I}_{\text{filt}})$ \\    
    
    \tcp{\textbf{Image Prompt Learning Module (Trainable: CATF, $\mathbf{P}$, MLP)}}
    $\mathbf{X}_{\text{fused}} \leftarrow \text{CATF}(\mathbf{X}_{\text{orig}}, \mathbf{X}_{\text{filt}})$ 
    \tcp*[f]{Instance-level prompting} \\
    $\mathbf{X}_{\text{VIT}} \leftarrow [\mathbf{x}_{\text{cls}}; \mathbf{P}; \mathbf{X}_{\text{fused}}]$ 
    \tcp*[f]{Shared-level prompting} \\
    $\mathbf{z}_I^{\text{VIT}} \leftarrow f_V(\mathbf{X}_{\text{VIT}})_{cls}$ \tcp*[f]{VIT with prompt guidance} \\
    $\mathbf{z}_I \leftarrow \text{MLP}(\mathbf{z}_I^{\text{VIT}})$ \tcp*[f]{Projection head maps image features to EEG space} \\

    \tcp{\textbf{Compute Cross-Modal Alignment Loss}}
    $\mathcal{L}_{\text{clip}} \leftarrow$ InfoNCE loss with $\mathbf{z}_E$ and $\mathbf{z}_I$ \\
    $\mathcal{L}_{\text{soft}} \leftarrow$ KL-based loss with intra-modality structure \\
    $\mathcal{L}_{\text{rel}} \leftarrow$ Relation-aware loss on negatives \\
    $\mathcal{L} \leftarrow \mu \mathcal{L}_{\text{clip}} + \alpha \mathcal{L}_{\text{soft}} + \lambda \mathcal{L}_{\text{rel}}$ 
    \tcp*[f]{Total training loss} \\

    \tcp{Parameter Update}
    Update $\theta_E, \theta_W, \theta_B, \theta_\phi, \theta_{\text{CATF}}, \theta_P, \theta_{\text{MLP}}$ using $\nabla \mathcal{L}$
}
\Return{$f_E$, $(\mathbf{W}, \mathbf{B})$, $\phi$, $\text{CATF}$, $\mathbf{P}$, $\text{MLP}$}
\end{algorithm}

\section{Experiments}

\subsection{Dataset}

In this study, we primarily conduct training and evaluation on the \textbf{THINGS-EEG2} dataset~\cite{gifford2022large}. To further assess the generalization ability of our model across different neural modalities, we also perform experiments on the THINGS-MEG dataset. Both datasets are constructed based on the THINGS image database and provide high-quality neural recordings aligned with visual semantic labels, making them well-suited for brain-vision alignment and cross-modal learning tasks.

The \textbf{THINGS-EEG2} dataset contains EEG recordings from 10 participants under the Rapid Serial Visual Presentation (RSVP) paradigm. The training set consists of 1,654 object concepts, each associated with 10 distinct images, and each image is presented four times per subject. The test set comprises 200 unseen concepts, with one image per concept repeated 80 times to enhance signal stability. Following the preprocessing protocol in ~\cite{wu2025bridging}, repetitions of the same stimulus are averaged to improve the signal-to-noise ratio (SNR). After preprocessing, each subject yields 16,540 training samples and 200 test samples.

The \textbf{THINGS-MEG} dataset includes Magnetoencephalography (MEG) recordings from 4 participants using a 271-channel whole-head MEG system. The training set contains 1,854 object concepts, each associated with 12 images shown once. The test set comprises 200 novel concepts, each with one image repeated 12 times. We adopt the same preprocessing strategy as used for the EEG data, averaging repeated trials to enhance SNR, and follow the experimental setup described in ~\cite{wu2025bridging} to ensure consistency.

\subsection{Encoder Configurations}

\textbf{EEG Encoders}

To evaluate the compatibility and alignment capability of our proposed fine-tuned CLIP-VIT image encoder with neural representations derived from EEG signals, we adopted several mainstream EEG encoders(including \textbf{EEGProject ~\cite{wu2025bridging}, TSconv ~\cite{song2023decoding}, Shallownet ~\cite{schirrmeister2017deep}, Deepnet ~\cite{schirrmeister2017deep}, EEGnet ~\cite{lawhern2018eegnet}}) that have been widely utilized and benchmarked in previous EEG-based decoding and classification studies. These include representative architectures from prior works, ensuring fair and standardized comparisons with existing baselines. Additionally, the \textbf{EEGFuseNet} ~\cite{liang2021eegfusenet} encoder was included in our experiments to assess its suitability for the task.

\textbf{Image Encoders}

In our experiments, we fine-tuned four versions of the CLIP-VIT model, namely \textbf{VIT-B/16, VIT-B/32, VIT-L/14, and VIT-H/14.} These models were obtained from the OpenCLIP ~\cite{ilharco_gabriel_2021_5143773} repository, which provides publicly available pretrained weights. To maintain the integrity of the original pretrained representations, the VIT backbones were kept frozen during training. Only the additional lightweight parameters introduced for our fine-tuning strategy were optimized. This setup allows us to examine the scalability and adaptability of our method across different model capacities while preserving computational efficiency.

\subsection{Implementation Details}

All experiments were implemented using PyTorch and conducted under Python versions 3.8, 3.9, and 3.10. The training and evaluation processes were distributed across three types of NVIDIA GPUs: A100 (40 GB), A40 (48 GB), and H20 (98 GB). To accommodate GPU memory limitations, different batch sizes were adopted based on the specific VIT model and GPU device, as detailed in Table~\ref{tab:batchsize}. Data preprocessing and loading strictly followed the protocol described in UBP~\cite{wu2025bridging}, ensuring consistency in data handling and partitioning. From the 16,540 training samples, 740 were randomly held out as a validation set. During training, the model checkpoint with the lowest validation loss was selected for final testing on 200 zero-shot samples. These 200 test samples belong to entirely unseen images, with no class overlap with the training or validation set, strictly conforming to the zero-shot retrieval setting widely used in the literature.

Our proposed model adopts a fully end-to-end training pipeline. For visual features, the pretrained CLIP-VIT backbones were frozen, and only the lightweight modules we introduced—namely, the dynamic filter generator, the distribution-aware fusion module, the visual prompt tuning layer, and the projection layer following CLIP-VIT—were updated during training. On the EEG side, all parameters in the EEG encoder and EEG perturbation module were set as trainable. Unless otherwise stated, we fixed the EEG input length to 1000 ms (downsampled to 250 Hz, resulting in 250 time points per sample) and selected 17 EEG channels located in the occipital and parietal regions, following the same setup as UBP. We also conducted separate ablation studies on the temporal and channel dimensions.

Each experiment was independently repeated ten times under identical conditions, and the reported results represent the average performance. For optimization, we adopted a dual-optimizer strategy. The first optimizer was used to update the EEG perturbation layer, EEG encoder, and the projection layer, with a learning rate of 0.002. The second optimizer was responsible for training the dynamic filter module, distribution-aware fusion module, and visual prompt tokens, using a learning rate of 0.02. The composite loss function was configured with weighting coefficients $\mu = 0.6$, $\alpha = 0.3$, and $\lambda = 0.1$, while the soft target coefficient was set to $\beta = 0.3$.

%BS SETTINGS
\newcolumntype{C}{>{\centering\arraybackslash}X}
\begin{table*}[htbp]
\centering
\renewcommand{\arraystretch}{1.3}
\footnotesize % 字体稍大于 footnotesize
\caption{Batch Size Settings Across Different GPUs and CLIP-VIT Variants}
\label{tab:batchsize}
\begin{tabularx}{\textwidth}{lCCC}
\toprule
\textbf{VIT Variant} & \textbf{A40 (48G)} & \textbf{A100 (40G)} & \textbf{H20 (98G)} \\
\midrule
VIT-B/16 & 128 & 128 & 256 \\
VIT-B/32 & 128 & 128 & 256 \\
VIT-L/14 & 128 & 64  & 256 \\
VIT-H/14 & 64  & 64  & 128 \\
\bottomrule
\end{tabularx}
\end{table*}

\subsection{Evaluation Metrics}
To quantitatively evaluate the effectiveness of our EEG-image alignment framework, we adopt three key evaluation metrics: Top-$k$ accuracy, mean Average Precision (mAP), and paired similarity score. These metrics comprehensively reflect the model’s retrieval performance and cross-modal representational alignment quality.

\textbf{Top-$k$ Accuracy} assesses whether the ground-truth image corresponding to an EEG query appears within the top $k$ most similar images retrieved based on the learned embeddings. Let $\mathcal{R}_i^{(k)}$ denote the top-$k$ retrieval set for the $i$-th EEG sample $\mathbf{e}_i$, and let $n$ denote the total number of test samples. The Top-$k$ accuracy is defined as:
\begin{equation}
\text{Top-}k = \frac{1}{n} \sum_{i=1}^n \mathbb{1}\{i \in \mathcal{R}_i^{(k)}\},
\end{equation}
where $\mathbb{1}\{\cdot\}$ is the indicator function returning 1 if the ground-truth image is ranked within the top-$k$ results.

\textbf{Mean Average Precision (mAP)} evaluates the quality of ranked retrieval results across all EEG queries. For a given query $\mathbf{e}_i$, let $\text{rel}_i(j) \in \{0,1\}$ denote whether the $j$-th retrieved image is relevant, and let the precision at rank $j$ be:
\begin{equation}
P_i(j) = \frac{1}{j} \sum_{l=1}^{j} \text{rel}_i(l).
\end{equation}
Then the mean Average Precision is computed as:
\begin{equation}
\text{mAP} = \frac{1}{n} \sum_{i=1}^n \left( \sum_{j=1}^{n} P_i(j) \cdot \text{rel}_i(j) \right).
\end{equation}

\textbf{Similarity Score} quantifies the alignment consistency between EEG and image representations in the joint embedding space. Given EEG features $\mathbf{e}_i$ and image features $\mathbf{v}_j$, both $\ell_2$-normalized, the cosine similarity between sample $i$ and $j$ is computed as:
\begin{equation}
\text{sim}(\mathbf{e}_i, \mathbf{v}_j) = \mathbf{e}_i^\top \mathbf{v}_j.
\label{eq:similatiryscorediv}
\end{equation}
This defines the similarity matrix $\mathbf{S}$:
\begin{equation}
\mathbf{S} = \mathbf{E} \cdot \mathbf{V}^\top \in \mathbb{R}^{n \times n}
\label{eq:similatiryscoreall}
\end{equation}
% To evaluate alignment on matched pairs only, we compute the average of the diagonal elements:
% \begin{equation}
% \text{SimScore} = \frac{1}{n} \sum_{i=1}^n \mathbf{S}_{i,i}.
% \end{equation}

\subsection{Performance Comparison}

To ensure fair and consistent benchmarking on the THINGS-EEG2 dataset, we follow the exact same evaluation protocol as proposed in UBP~\cite{wu2025bridging}. We compare our method against several state-of-the-art baselines, including BraVL~\cite{du2023decoding}, NICE~\cite{song2023decoding}, ATM-S~\cite{li2024visual}, VE-SDN~\cite{chen2024visual}, and UBP~\cite{wu2025bridging}. These methods represent a diverse set of EEG-image alignment strategies, such as multimodal mixture-of-experts, self-supervised learning, and semantic disentanglement.

In addition, on the THINGS-MEG dataset, we primarily compare our model with the NICE~\cite{song2023decoding} and UBP~\cite{wu2025bridging} frameworks, as these are among the few prior methods that have reported results under comparable settings on this benchmark. Both approaches serve as strong baselines, and our comparison is conducted under identical intra-subject and inter-subject evaluation settings to ensure result consistency.

Table~\ref{tab:THINGSEEGINTRA} reports the Top-1 and Top-5 accuracy across ten subjects on the THINGS-EEG2 dataset. Among all baselines, UBP~\cite{wu2025bridging} previously achieved the best performance with an average Top-1 of 50.9\% and Top-5 of 79.7\%. Our proposed NeuroCLIP significantly outperforms UBP, reaching 63.2\% Top-1 and 90.3\% Top-5 accuracy, with relative gains of +12.3\% and +10.6\%, respectively. NeuroCLIP consistently achieves higher scores across all subjects. For example, on Subject 2 and Subject 10, it reaches 64.5\% and 69.1\% Top-1 accuracy, outperforming UBP by a clear margin. Compared to earlier baselines such as NICE and VE-SDN, NeuroCLIP demonstrates substantial improvements, confirming the effectiveness of our proposed alignment strategy. 

\begin{figure}[htbp]
    \centering
    \includegraphics[width=\textwidth]{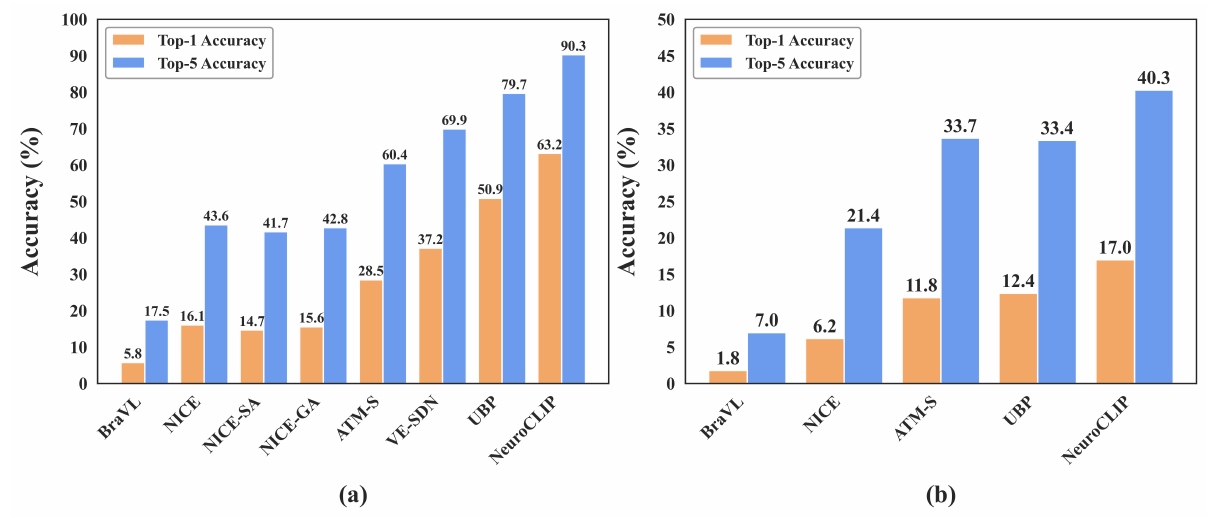}
    \caption{Comparison of average Top-1 and Top-5 accuracy across different methods under (a) intra-subject and (b) inter-subject settings on the THINGS-EEG2 dataset.}
    \label{fig:bar_comparison_combined}
\end{figure}

Table~\ref{tab:THINGSEEGINTER} summarizes the Top-1 and Top-5 accuracy under the inter-subject evaluation setting. Among the compared methods, UBP~\cite{wu2025bridging} previously achieved the highest average Top-1 accuracy of 12.4\% and Top-5 accuracy of 33.4\%. Our proposed NeuroCLIP significantly outperforms UBP in this more challenging cross-subject scenario, achieving an average Top-1 accuracy of 17.0\% and Top-5 accuracy of 40.3\%. This represents a clear relative improvement of +4.6\% in Top-1 and +6.9\% in Top-5 performance. Notably, NeuroCLIP exhibits substantial gains on several subjects, such as Subject 2 (Top-1: 31.8\% vs. 15.5\%) and Subject 10 (Top-1: 29.8\% vs. 16.0\%), highlighting its stronger generalization ability across individuals. Compared to earlier methods like NICE and ATM-S, which show more limited performance, NeuroCLIP consistently achieves superior results. These findings further confirm the robustness of our alignment strategy under inter-subject conditions, where EEG signal variability is higher and model generalization is more critical. To provide a clearer comparison of model performance on the THINGS-EEG2 dataset, we visualize the average Top-1 and Top-5 200-way retrieval accuracy under both intra-subject and inter-subject settings, as illustrated in Figure~\ref{fig:bar_comparison_combined}. Our NeuroCLIP consistently outperforms all baselines in both settings, demonstrating its superior cross-modal alignment capabilities.

%THINGEEGintra
\begin{table}[htbp]
\centering
\caption{Comparison of Top-1 and Top-5 accuracy on the THINGS-EEG2 dataset(intra-subject)}
\label{tab:THINGSEEGINTRA}
\renewcommand{\arraystretch}{1.3} % 控制行高
\large % 整体字号放大两号
\setlength{\tabcolsep}{2pt}
\resizebox{\textwidth}{!}{%
\begin{tabular}{c*{11}{cc}}
\toprule
\multirow{2}{*}{Method} &
\multicolumn{2}{c}{Subject1} &
\multicolumn{2}{c}{Subject2} &
\multicolumn{2}{c}{Subject3} &
\multicolumn{2}{c}{Subject4} &
\multicolumn{2}{c}{Subject5} &
\multicolumn{2}{c}{Subject6} &
\multicolumn{2}{c}{Subject7} &
\multicolumn{2}{c}{Subject8} &
\multicolumn{2}{c}{Subject9} &
\multicolumn{2}{c}{Subject10} &
\multicolumn{2}{c}{Avg} \\
\cmidrule(lr){2-3}  \cmidrule(lr){4-5}  \cmidrule(lr){6-7}
\cmidrule(lr){8-9}  \cmidrule(lr){10-11} \cmidrule(lr){12-13}
\cmidrule(lr){14-15} \cmidrule(lr){16-17} \cmidrule(lr){18-19}
\cmidrule(lr){20-21} \cmidrule(lr){22-23}
\footnotesize & Top-1 & Top-5 & Top-1 & Top-5 & Top-1 & Top-5 & Top-1 & Top-5
& Top-1 & Top-5 & Top-1 & Top-5 & Top-1 & Top-5 & Top-1 & Top-5
& Top-1 & Top-5 & Top-1 & Top-5 & Top-1 & Top-5 \\
\midrule
BraVL\cite{du2023decoding} & 6.1 & 17.9 & 4.9 & 14.9 & 5.6 & 17.4 & 5.0 & 15.1 & 4.0 & 13.4 & 6.0 & 18.2 & 6.5 & 20.4 & 8.8 & 23.7 & 4.3 & 14.0 & 7.0 & 19.7 & 5.8 & 17.5 \\
NICE\cite{song2023decoding} & 13.2 & 39.5 & 13.5 & 40.3 & 14.5 & 42.7 & 20.6 & 52.7 & 10.1 & 31.5 & 16.5 & 44.0 & 17.0 & 42.1 & 22.9 & 56.1 & 15.4 & 41.6 & 17.4 & 45.8 & 16.1 & 43.6 \\
NICE-SA\cite{song2023decoding} & 13.3 & 40.2 & 12.1 & 36.1 & 15.3 & 39.6 & 15.9 & 49.0 & 9.8 & 34.4 & 14.2 & 42.4 & 17.9 & 43.6 & 18.2 & 50.2 & 14.4 & 38.7 & 16.0 & 42.8 & 14.7 & 41.7 \\
NICE-GA\cite{song2023decoding} &15.2 & 40.1 & 13.9 & 40.1 & 14.7 & 42.7 & 17.6 & 48.9 & 9.0 & 29.7 & 16.4 & 44.4 & 14.9 & 43.1 & 20.3 & 52.1 & 14.1 & 39.7 & 19.6 & 46.7 & 15.6 & 42.8\\
ATM-S\cite{li2024visual} & 25.6  & 60.4 & 22.0 & 54.5 & 25.0 & 62.4 & 31.4 & 60.9
& 12.9 & 43.0 & 21.3 & 51.1 & 30.5 & 61.5 & 38.8 & 72.0
& 34.4 & 51.5 & 29.1 & 63.5 & 28.5 & 60.4 \\
VE-SDN\cite{chen2024visual}   &32.6 & 63.7 & 34.4 & 69.9 & 38.7 & 73.5 & 39.8 & 72.0 & 29.4 & 58.6 & 34.5 & 68.8 & 34.5 & 68.3 & 49.3 & 79.8 & 39.0 & 69.6 & 39.8 & 75.3 & 37.2 & 69.9 \\
UBP\cite{wu2025bridging} &41.2 & 70.5 & 51.2 & 80.9 & 51.2 & 82.0 & 51.1 & 76.9 & 42.2 & 72.8 & 57.5 & 83.5 & 49.0 & 79.9 & 58.6 & 85.8 & 45.1 & 76.2 & 61.5 & 88.2 & 50.9 & 79.7 \\
\rowcolor{gray!15}
\textbf{NeuroCLIP(ours)}   &60.3 & 84.2 & 64.5 & 94.1 & 63.7 & 93.2 & 62.8 & 91.3 & 55.0 & 83.7 & 68.6 & 93.6 & 70.6 & 93.3 & 69.6 & 90.3 & 48.2 & 86.5 & 69.1 & 92.1 & \textbf{63.2} & \textbf{90.3}\\
\bottomrule
\end{tabular}%
}
\end{table}

%THINGSEEGinter
\begin{table}[htbp]
\centering
\caption{Comparison of Top-1 and Top-5 accuracy on the THINGS-EEG2 dataset(inter-subject)}
\label{tab:THINGSEEGINTER}
\renewcommand{\arraystretch}{1.3} % 控制行高
\large % 整体字号放大两号
\setlength{\tabcolsep}{2pt}
\resizebox{\textwidth}{!}{%
\begin{tabular}{c*{11}{cc}}
\toprule
\multirow{2}{*}{Method} &
\multicolumn{2}{c}{Subject1} &
\multicolumn{2}{c}{Subject2} &
\multicolumn{2}{c}{Subject3} &
\multicolumn{2}{c}{Subject4} &
\multicolumn{2}{c}{Subject5} &
\multicolumn{2}{c}{Subject6} &
\multicolumn{2}{c}{Subject7} &
\multicolumn{2}{c}{Subject8} &
\multicolumn{2}{c}{Subject9} &
\multicolumn{2}{c}{Subject10} &
\multicolumn{2}{c}{Avg} \\
\cmidrule(lr){2-3}  \cmidrule(lr){4-5}  \cmidrule(lr){6-7}
\cmidrule(lr){8-9}  \cmidrule(lr){10-11} \cmidrule(lr){12-13}
\cmidrule(lr){14-15} \cmidrule(lr){16-17} \cmidrule(lr){18-19}
\cmidrule(lr){20-21} \cmidrule(lr){22-23}
\footnotesize & Top-1 & Top-5 & Top-1 & Top-5 & Top-1 & Top-5 & Top-1 & Top-5
& Top-1 & Top-5 & Top-1 & Top-5 & Top-1 & Top-5 & Top-1 & Top-5
& Top-1 & Top-5 & Top-1 & Top-5 & Top-1 & Top-5 \\
\midrule
BraVL\cite{du2023decoding} & 2.3 & 8.0 & 1.5 & 6.3 & 1.4 & 5.9 & 1.7 & 6.7 & 1.5 & 5.6 & 1.8 & 7.2 & 2.1 & 8.1 & 2.2 & 7.6 & 1.6 & 6.4 & 2.3 & 8.5 & 1.8 & 7.0 \\
NICE\cite{song2023decoding} & 7.6 & 22.8 & 5.9 & 20.5 & 6.0 & 22.3 & 6.3 & 20.7 & 4.4 & 18.3 & 5.6 & 22.2 & 5.6 & 19.7 & 6.3 & 22.0 & 5.7 & 17.6 & 8.4 & 28.3 & 6.2 & 21.4 \\
ATM-S\cite{li2024visual} & 10.5 & 26.8 & 7.1 & 24.8 & 11.9 & 33.8 & 14.7 & 39.4 & 7.0 & 23.9 & 11.1 & 35.8 & 16.1 & 43.5 & 15.0 & 40.3 & 4.9 & 22.7 & 20.5 & 46.5 & 11.8 & 33.7 \\
UBP\cite{wu2025bridging} & 11.5 & 29.7 & 15.5 & 40.0 & 9.8 & 27.0 & 13.0 & 32.3 & 8.8 & 33.8 & 11.7 & 31.0 & 10.2 & 23.8 & 12.2 & 32.2 & 15.5 & 40.5 & 16.0 & 43.5 & 12.4 & 33.4 \\
\rowcolor{gray!15}
\textbf{NeuroCLIP(ours)}   &17.2 & 47.3 & 31.8 & 60.2 & 10.9 & 24.3 & 19.1 & 48.2 & 14.3 & 32.6 & 11.2 & 32.1 & 15.8 & 41.3 & 10.2 & 29.2 & 9.9 & 27.4 & 29.8 & 60.0 &\textbf{17.0} &\textbf{40.3} \\
\bottomrule
\end{tabular}%
}
\end{table}

To further evaluate the generalizability of our model, we conduct additional experiments on the THINGS-MEG dataset. As shown in Table~\ref{tab:THINGSMEG_ALL}, NeuroCLIP achieves competitive performance under both intra- and inter-subject settings. Notably, our method demonstrates strong generalization on Subject2 and Subject3 in the intra-subject scenario and shows consistent improvements over the UBP baseline in the inter-subject case, validating the robustness of our framework on more challenging MEG signals.
%Things-meg interandintra
\begin{table}[htbp]
\centering
\caption{Comparison of Top-1 and Top-5 accuracy on the THINGS-MEG dataset (intra-subject \& inter-subject)}
\label{tab:THINGSMEG_ALL}
\renewcommand{\arraystretch}{1.2}
\resizebox{\textwidth}{!}{%
  {\scriptsize
  \begin{tabular}{c*{5}{cc}}
  \toprule
  \multirow{2}{*}{Method} &
  \multicolumn{2}{c}{Subject1} &
  \multicolumn{2}{c}{Subject2} &
  \multicolumn{2}{c}{Subject3} &
  \multicolumn{2}{c}{Subject4} &
  \multicolumn{2}{c}{Avg} \\
  \cmidrule(lr){2-3} \cmidrule(lr){4-5} \cmidrule(lr){6-7}
  \cmidrule(lr){8-9} \cmidrule(lr){10-11}
  \footnotesize & Top-1 & Top-5 & Top-1 & Top-5 & Top-1 & Top-5 & Top-1 & Top-5 & Top-1 & Top-5 \\
  \midrule
  \multicolumn{11}{c}{\textbf{Intra-subject}} \\
  \midrule
  NICE\cite{song2023decoding} & 9.6 & 27.8 & 18.5 & 47.8 & 14.2 & 41.6 & 9.0 & 26.6 & 12.8 & 36.0 \\
  NICE-SA\cite{song2023decoding} & 9.8 & 27.8 & 18.6 & 46.4 & 10.5 & 38.4 & 11.7 & 27.2 & 12.7 & 35.0 \\
  NICE-GA\cite{song2023decoding} & 8.7 & 30.5 & 21.8 & 56.6 & 16.5 & 49.7 & 10.3 & 32.3 & 14.3 & 42.3 \\
  UBP\cite{wu2025bridging} & 15.0 & 38.0 & 46.0 & 80.5 & 27.3 & 59.0 & 18.5 & 43.5 & 26.7 & 55.2 \\
  \rowcolor{gray!15}
  \textbf{NeuroCLIP(ours)} & 15.6 & 31.7 & 67.1 & 91.6 & 35.3 & 74.1 & 12.3 & 28.7 & \textbf{32.6} & \textbf{56.5} \\
  \midrule
  \multicolumn{11}{c}{\textbf{Inter-subject}} \\
  \midrule
  UBP\cite{wu2025bridging} & 2.0 & 5.7 & 1.5 & 17.2 & 2.7 & 10.5 & 2.5 & 8.0 & 2.2 & 10.4 \\
  \rowcolor{gray!15}
  \textbf{NeuroCLIP(ours)} & 2.1 & 6.0 & 2.7 & 16.3 & 2.9 & 12.4 & 2.2 & 8.9 & \textbf{2.5} & \textbf{10.9} \\
  \bottomrule
  \end{tabular}
  }
}
\end{table}

%barchart cross methods

\section{Discussions}
\subsection{Ablation Studies}

\subsubsection{Ablation on Core Architectural Components}
We conduct comprehensive ablation studies under both intra-subject and inter-subject paradigms on the THINGS-EEG2 dataset to evaluate the individual contributions of key modules within the NeuroCLIP framework. Specifically, we consider the following ablated variants:
\begin{itemize}
    \item \textbf{w/o Dual-stream}: This variant removes the dual-stream image encoding strategy and retains the standard single-stream visual encoder as used in the original CLIP. The dynamic filtering mechanism is disabled, and only the raw image is embedded into patch-level features. As a result, no instance-level prompt token is generated. The model relies solely on shared-level prompts, without any bidirectional interaction between visual features and prompts.
    
    \item \textbf{w/o CABF}: In this setting, we disable our proposed cross-attention based token-level fusion strategy. Instead, the bilinear adapter fusion baseline (as formulated in Equation~\ref{eq:filtered_mix}) is adopted.

    \item \textbf{w/o Shared-level Prompt}: We remove the shared-level prompt token and retain only the instance-level prompts generated from filtered visual patches.

    \item \textbf{w/o EEG Perturbation}: The learnable linear perturbation module applied to EEG embeddings is removed, meaning the EEG encoder operates without the proposed perturbation-based regularization.

    \item \textbf{w/o Soft Target Formulation}: The improved soft target formulation in our contrastive loss is replaced with the standard InfoNCE loss used in conventional CLIP training.
\end{itemize}
The experimental results are summarized in Tables~\ref{tab:intra_ablation} and~\ref{tab:inter_ablation}.

In the intra-subject setting, removing any single component leads to a noticeable drop in performance, underscoring the effectiveness of each module. Notably, the visual prompt token mechanism shows the greatest impact, with a significant reduction in average Top-1 accuracy from 63.2\% to 44.5\%, demonstrating its critical role in enhancing visual-semantic alignment. Similarly, components such as the dynamic filter, EEG perturbation, and soft contrastive loss each contribute to improving both discriminative capacity and modality alignment, as evidenced by performance declines when they are removed.

In the more challenging inter-subject scenario, the ablation results reveal a similar trend. Among the modules, the dynamic filtering and EEG perturbation components are particularly vital for mitigating inter-subject variability and preserving semantic consistency. Even in the absence of visual prompt tokens, the performance drops sharply to 12.0\% (Top-1), indicating the importance of task-adaptive prompting in handling subject-level variations.

%Intra-ablation
\begin{table}[htbp]
\centering
\caption{Ablation study on THINGS-EEG2 dataset(intra-subject)}
\label{tab:intra_ablation}
\renewcommand{\arraystretch}{1.3}
\large
\setlength{\tabcolsep}{2pt}
\resizebox{\textwidth}{!}{%
\begin{tabular}{c*{11}{cc}}
\toprule
\multirow{2}{*}{Method} &
\multicolumn{2}{c}{Subject1} &
\multicolumn{2}{c}{Subject2} &
\multicolumn{2}{c}{Subject3} &
\multicolumn{2}{c}{Subject4} &
\multicolumn{2}{c}{Subject5} &
\multicolumn{2}{c}{Subject6} &
\multicolumn{2}{c}{Subject7} &
\multicolumn{2}{c}{Subject8} &
\multicolumn{2}{c}{Subject9} &
\multicolumn{2}{c}{Subject10} &
\multicolumn{2}{c}{Avg} \\
\cmidrule(lr){2-3}  \cmidrule(lr){4-5}  \cmidrule(lr){6-7}
\cmidrule(lr){8-9}  \cmidrule(lr){10-11} \cmidrule(lr){12-13}
\cmidrule(lr){14-15} \cmidrule(lr){16-17} \cmidrule(lr){18-19}
\cmidrule(lr){20-21} \cmidrule(lr){22-23}
\footnotesize & Top-1 & Top-5 & Top-1 & Top-5 & Top-1 & Top-5 & Top-1 & Top-5
& Top-1 & Top-5 & Top-1 & Top-5 & Top-1 & Top-5 & Top-1 & Top-5
& Top-1 & Top-5 & Top-1 & Top-5 & Top-1 & Top-5 \\
\midrule
w/o Dual-stream & 46.2 & 77.8 & 55.6 & 82.0 & 59.0 & 79.5 & 48.3 & 80.1 & 41.5 & 72.9 & 57.4 & 80.9 & 51.5 & 77.6 & 60.2 & 85.2 & 43.0 & 74.7 & 55.2 & 81.1 & 51.8 & 79.2 \\
w/o CABF & 55.1 & 80.8 & 63.1 & 89.3 & 49.9 & 79.0 & 51.5 & 82.6 & 59.5 & 88.0 & 53.0 & 86.2 & 57.8 & 89.1 & 60.0 & 90.2 & 53.3 & 85.1 & 59.6 & 88.2 & 56.3 & 85.9 \\
w/o Shared-level Prompt & 42.7 & 71.5 & 40.4 & 69.5 & 55.6 & 82.7 & 46.0 & 73.6 & 31.5 & 62.0 & 41.5 & 69.9 & 47.3 & 73.2 & 49.2 & 80.5 & 38.5 & 68.5 & 52.6 & 81.5 & 44.5 & 73.3 \\
w/o EEG Perturbation & 51.3 & 79.7 & 56.9 & 90.0 & 59.8 & 87.1 & 63.0 & 92.1 & 38.9 & 65.6 & 61.7 & 91.2 & 50.8 & 83.1 & 60.5 & 90.8 & 52.5 & 80.0 & 62.7 & 93.1 & 55.8 & 85.3 \\
w/o Soft Target Formulation & 47.5 & 85.7 & 49.3 & 86.2 & 51.8 & 89.0 & 43.0 & 72.3 & 29.1 & 60.5 & 46.0 & 81.0 & 40.2 & 77.8 & 52.2 & 89.0 & 40.5 & 81.5 & 53.4 & 85.9 & 45.4 & 80.9 \\
\rowcolor{gray!15}
\textbf{NeuroCLIP} &60.3 & 84.2 & 64.5 & 94.1 & 63.7 & 93.2 & 62.8 & 91.3 & 55.0 & 83.7 & 68.6 & 93.6 & 70.6 & 93.3 & 69.6 & 90.3 & 48.2 & 86.5 & 69.1 & 92.1 & \textbf{63.2} & \textbf{90.3}\\
\bottomrule
\end{tabular}%
}
\end{table}

%Inter-ablation
\begin{table}[htbp]
\centering
\caption{Ablation study on THINGS-EEG2 dataset(inter-subject)}
\label{tab:inter_ablation}
\renewcommand{\arraystretch}{1.3} % 控制行高
\large % 整体字号放大两号
\setlength{\tabcolsep}{2pt}
\resizebox{\textwidth}{!}{%
\begin{tabular}{c*{11}{cc}}
\toprule
\multirow{2}{*}{Method} &
\multicolumn{2}{c}{Subject1} &
\multicolumn{2}{c}{Subject2} &
\multicolumn{2}{c}{Subject3} &
\multicolumn{2}{c}{Subject4} &
\multicolumn{2}{c}{Subject5} &
\multicolumn{2}{c}{Subject6} &
\multicolumn{2}{c}{Subject7} &
\multicolumn{2}{c}{Subject8} &
\multicolumn{2}{c}{Subject9} &
\multicolumn{2}{c}{Subject10} &
\multicolumn{2}{c}{Avg} \\
\cmidrule(lr){2-3}  \cmidrule(lr){4-5}  \cmidrule(lr){6-7}
\cmidrule(lr){8-9}  \cmidrule(lr){10-11} \cmidrule(lr){12-13}
\cmidrule(lr){14-15} \cmidrule(lr){16-17} \cmidrule(lr){18-19}
\cmidrule(lr){20-21} \cmidrule(lr){22-23}
\footnotesize & Top-1 & Top-5 & Top-1 & Top-5 & Top-1 & Top-5 & Top-1 & Top-5
& Top-1 & Top-5 & Top-1 & Top-5 & Top-1 & Top-5 & Top-1 & Top-5
& Top-1 & Top-5 & Top-1 & Top-5 & Top-1 & Top-5 \\
\midrule % 新增长横线
w/o Dual-stream & 8.6 & 20.9 & 21.5 & 40.9 & 6.3 & 14.1 & 12.3 & 40.2 & 5.8 & 21.9 & 7.7 & 26.7 & 10.1 & 30.8 & 5.1 & 18.3 & 4.2 & 13.8 & 16.6 & 39.9 & 9.8 & 26.8 \\
w/o CABF & 9.3 & 21.7 & 17.3 & 31.4 & 9.0 & 17.3 & 15.0 & 46.9 & 6.0 & 19.7 & 3.1 & 13.5 & 20.2 & 47.4 & 6.0 & 18.9 & 6.6 & 16.4 & 19.2 & 40.0 & 11.2 & 27.3 \\
w/o Shared-level Prompt & 5.5 & 10.2 & 13.3 & 27.1 & 3.5 & 9.2 & 12.6 & 26.3 & 10.6 & 27.4 & 9.0 & 22.5 & 19.3 & 26.0 & 8.8 & 21.6 & 13.5 & 31.0 & 24.1 & 53.5 & 12.0 & 25.5 \\
w/o EEG Perturbation & 16.1 & 42.3 & 33.4 & 50.1 & 8.2 & 20.1 & 16.7 & 46.1 & 15.2 & 36.1 & 9.0 & 30.0 & 17.5 & 40.9 & 17.7 & 40.2 & 6.5 & 19.9 & 36.6 & 47.4 & 17.7 & 37.3 \\
w/o Soft Target Formulation & 13.5 & 38.7 & 25.0 & 46.1 & 6.8 & 17.3 & 12.4 & 33.1 & 12.8 & 31.8 & 9.7 & 29.9 & 9.3 & 29.8 & 9.2 & 31.3 & 5.4 & 20.5 & 21.5 & 46.7 & 12.6 & 32.5 \\
\rowcolor{gray!15}
\textbf{NeuroCLIP}   &17.2 & 47.3 & 31.8 & 60.2 & 10.9 & 24.3 & 19.1 & 48.2 & 14.3 & 32.6 & 11.2 & 32.1 & 15.8 & 41.3 & 10.2 & 29.2 & 9.9 & 27.4 & 29.8 & 60.0 &\textbf{17.0} &\textbf{40.3} \\
\bottomrule
\end{tabular}%
}
\end{table}

\subsubsection{Ablation on EEG Channel and Temporal Segments}
\begin{figure}[htbp]
    \centering
    \includegraphics[width=0.8\textwidth]{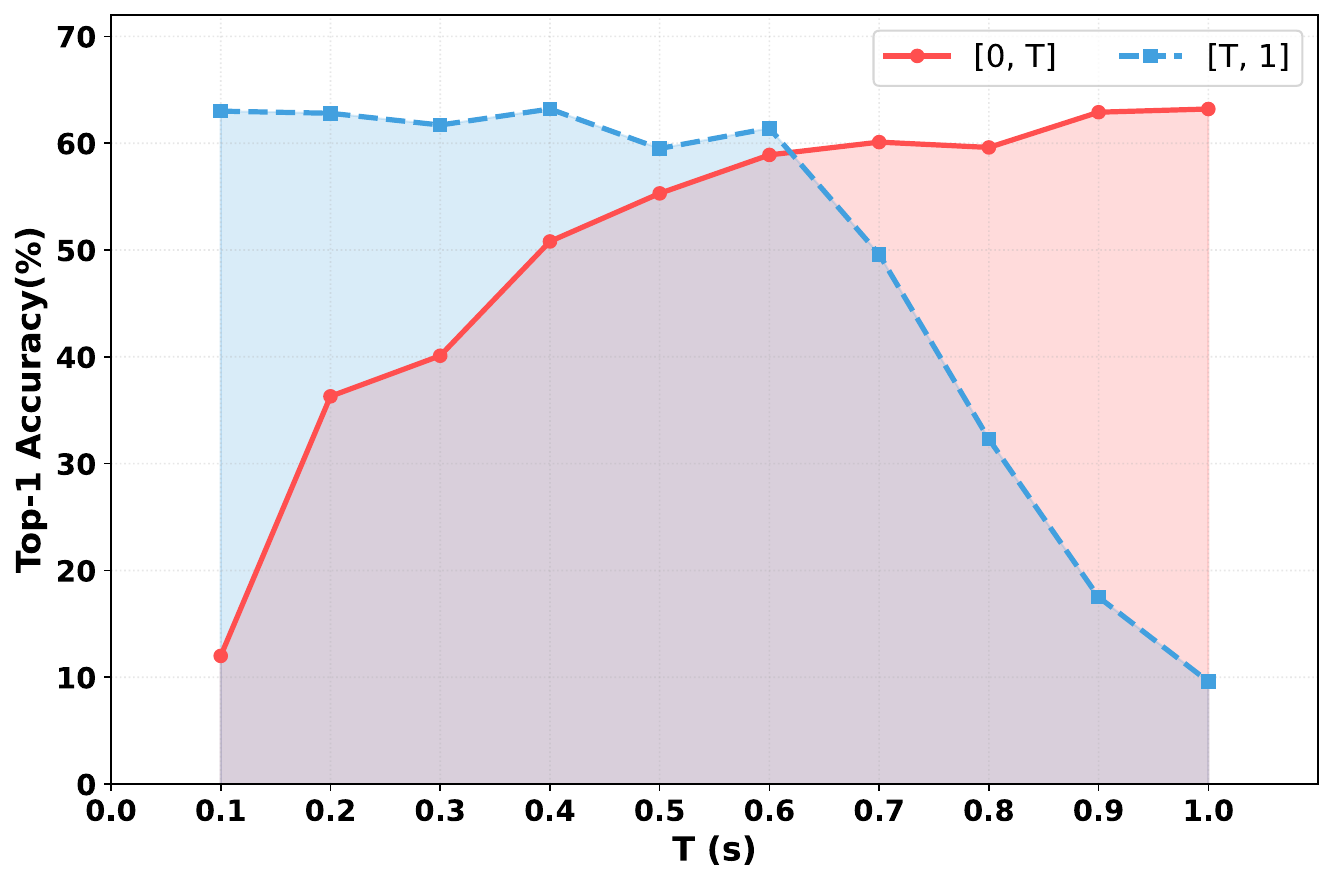}  % 推荐改文件名
    % 若不想改文件名，可保留括号，但建议加引号：\includegraphics[width=0.8\textwidth]{"accuracy_top1(2).pdf"}
    \caption{Ablation study on temporal segment}
    \label{fig:time_ablation}
\end{figure}

To further examine the impact of spatial and temporal EEG characteristics on visual decoding, we conducted ablation studies on both channel region and temporal window selection. The results are visualized in table~\ref{tab:channel_ablation} and Figure~\ref{fig:time_ablation}. 

From the channel-wise analysis (table~\ref{tab:channel_ablation}), we observe that using only occipital electrodes achieves Top-1 accuracy of 55.2\%, which already surpasses the result obtained from using all channels (51.6\%). This aligns with neuroscience findings that occipital regions are primarily responsible for visual processing. The optimal channel selection, composed of a subset across visual-relevant regions—further improves Top-1 and Top-5 accuracy to 63.2\% and 90.3\%, respectively, confirming that carefully selecting brain areas enhances discriminative signal quality.

Temporal analysis (Figure~\ref{fig:time_ablation}) investigates decoding performance as a function of EEG segment duration. The blue curve shows the model performance when using EEG data from time 0 to $T$, while the orange curve represents using data from $T$ to 1 second. Results indicate that early segments (0$ \sim $0.6s) progressively improve performance, with Top-1 accuracy peaking around 65\% when using the full 1-second interval. In contrast, using only the later segments (after $T$) leads to sharp performance degradation beyond 0.6s, indicating that early post-stimulus EEG segments carry more visual-relevant information.

%ablation_channel
% \begin{figure}[htbp]
%     \centering
%     \includegraphics[width=\textwidth]{EEG_Channel_Ablation_Corrected_Bold_AdjustedBar.pdf}
%     \caption{Ablation study on brain area}
%     \label{fig:channel_ablation}
% \end{figure}

%ablation_channel_table
\begin{table}[htbp]
\centering
\renewcommand{\arraystretch}{1.3}
\caption{Performance across Different EEG Channel Selections (Unit: \%)}
\label{tab:channel_ablation}
\resizebox{\textwidth}{!}{%
\begin{tabular}{lccccccc}
\toprule
\textbf{Metric} & \textbf{All} & \textbf{Frontal (F)} & \textbf{Central (C)} & \textbf{Temporal (T)} & \textbf{Parietal (P)} & \textbf{Occipital (O)} & \textbf{Optimal Selection} \\
\midrule
Top-1 & 51.6 & 1.4 & 1.3 & 7.4 & 23.0 & 55.2 & \textbf{63.2} \\
Top-3 & 76.4 & 4.2 & 4.1 & 17.1 & 41.2 & 78.6 & \textbf{84.6} \\
Top-5 & 85.3 & 7.0 & 6.4 & 24.4 & 51.9 & 85.9 & \textbf{90.3} \\
mAP   & 66.0 & 5.8 & 5.5 & 17.2 & 36.7 & 68.9 & \textbf{75.1} \\
\bottomrule
\end{tabular}
}
\end{table}

%ablation_temporal

% \begin{figure}[htbp]
%     \centering
%     \includesvg[width=0.8\textwidth]{accuracy_top1(2).pdf}
%     \caption{Ablation study on temporal segment}
%     \label{fig:time_ablation}
% \end{figure}

\subsection{Analysis of EEG and Vision Encoder Combinations}
\begin{figure}[htbp]
    \centering
    \includegraphics[width=\textwidth]{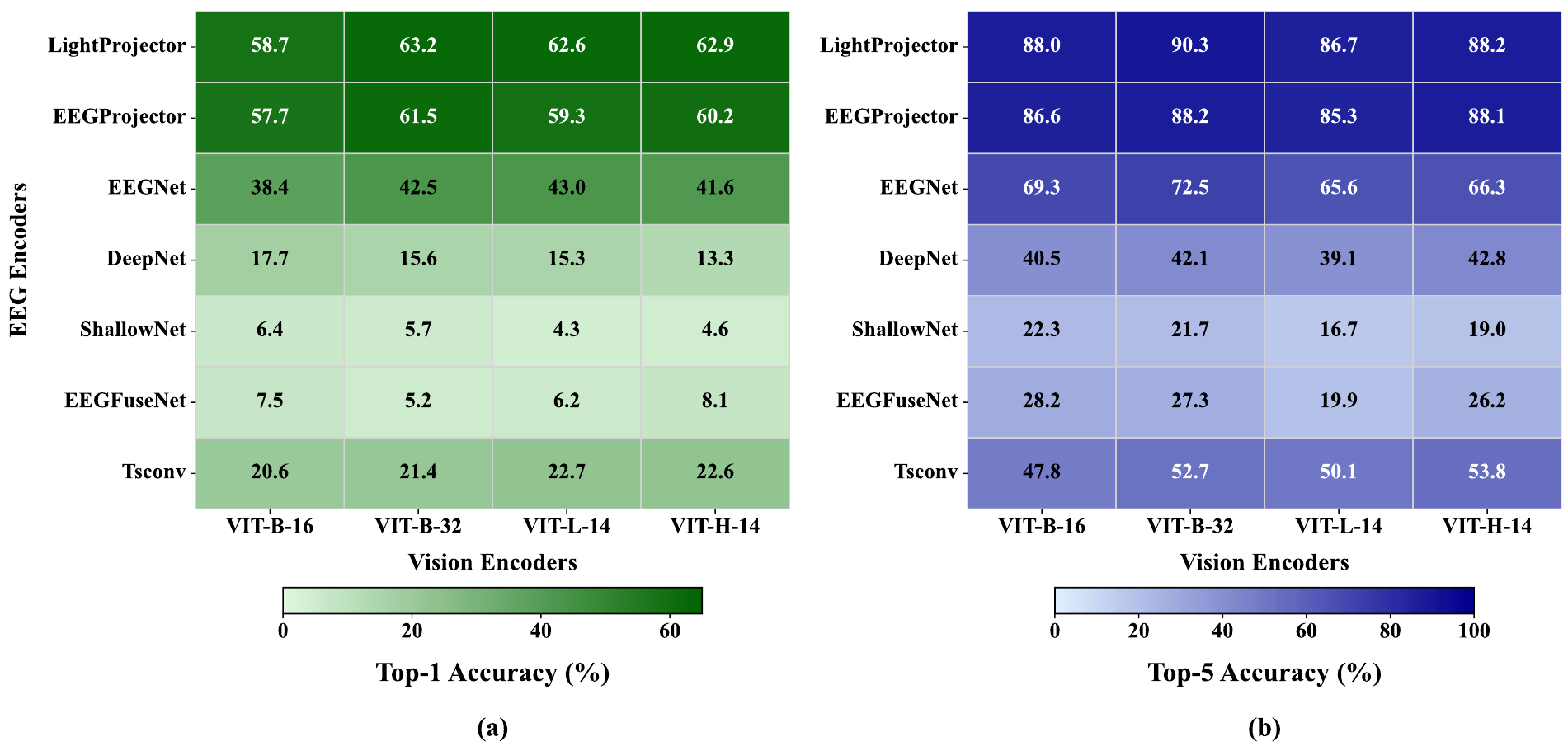}
    \caption{Visualization of different Encoders performance}
    \label{fig:eeg_vision_encoders}
\end{figure}

UBP \cite{wu2025bridging} previously proposed a lightweight and efficient EEG encoder, EEGProjector, which achieved impressive performance. In our study, benefiting from a more efficient alignment framework, we were able to further simplify the EEG encoder by proposing an even lighter variant, LightProjector. Specifically, we discarded the residual connection structure of EEGProjector and adopted a single fully connected layer to decode the EEG signals. This simplification reduced the number of parameters in the EEG encoder by half, while achieving even better alignment performance.
To explore the performance impact of different encoder pairings, we systematically evaluated seven EEG encoders in combination with four versions of the CLIP-VIT vision encoder. The results are presented in Figure~\ref{fig:eeg_vision_encoders}
%heatmapwithdiffEncoders
Among all combinations, LightProjector + VIT-B/32 achieves the best performance, with 63.2\% Top-1 and 90.3\% Top-5 accuracy. 

% In comparison, traditional EEG encoders like EEGNet, DeepNet, and ShallowNet yield lower accuracy, while TSconv performs reasonably well, especially in Top-5 accuracy. On the vision side, VIT-B/32 and VIT-H/14 generally lead to stronger results, suggesting the benefit of using mid-to-high capacity VIT backbones.

\subsection{Analysis of the Impact of Different Numbers of Prompt Tokens }
\begin{figure}[htbp]
    \centering
    \includegraphics[width=\textwidth]{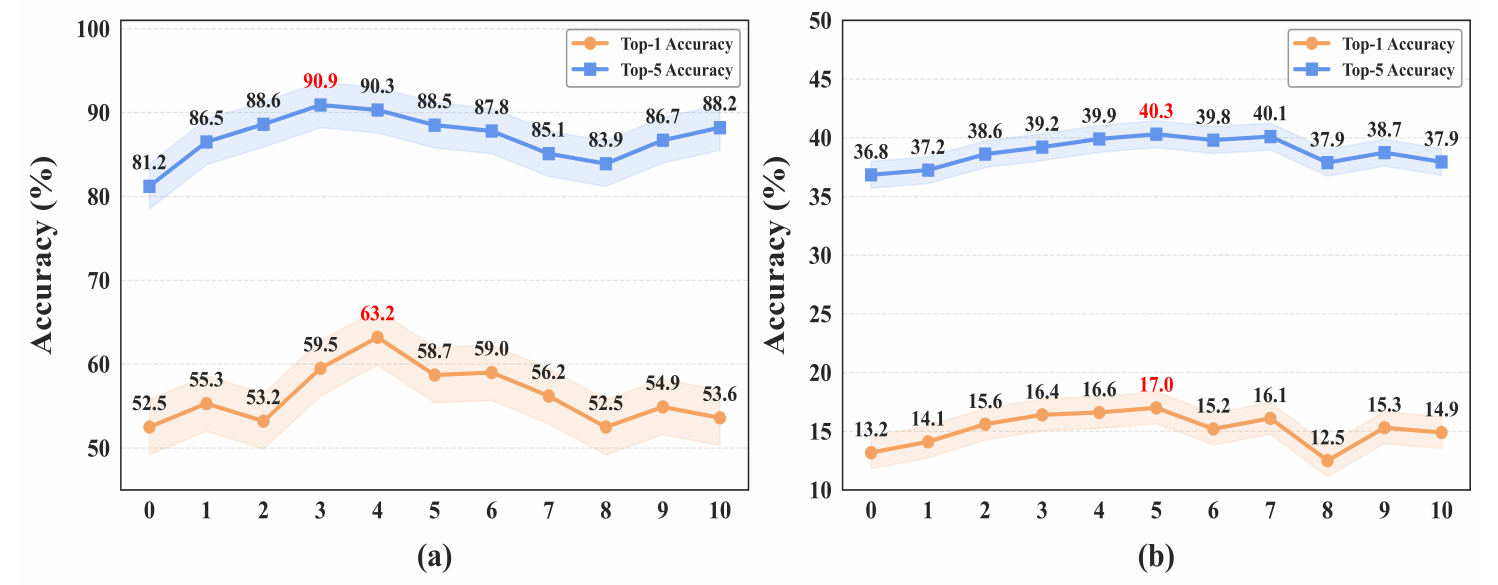}
    \caption{Comparison of average Top-1 and Top-5 accuracy across different numbers of visual prompt tokens under (a) intra-subject and (b) inter-subject settings.}
    \label{fig:prompt_token_analysis}
\end{figure}

Figure~\ref{fig:prompt_token_analysis} illustrates the impact of varying the number of visual prompt tokens on model performance under intra-subject and inter-subject settings on the THINGS-EEG2 dataset. In the intra-subject scenario, we observe that both Top-1 and Top-5 accuracy improve significantly as the number of prompt tokens increases, reaching optimal performance at around 4 tokens. Beyond this point, performance slightly declines, suggesting that excessive prompt tokens may introduce redundancy or noise.
The inter-subject setting shows optimal performance at 5 prompt tokens, indicating a slightly higher token count may benefit generalization across subjects. Beyond this point, accuracy begins to drop, likely due to increased variability and over-parameterization. These results confirm the importance of prompt token tuning for robust EEG-image alignment under varying generalization conditions.
%plottokennumbers

\subsection{Analysis of Sample Similarity Prediction}

\begin{figure}[htbp]
    \centering
    \includegraphics[width=\textwidth]{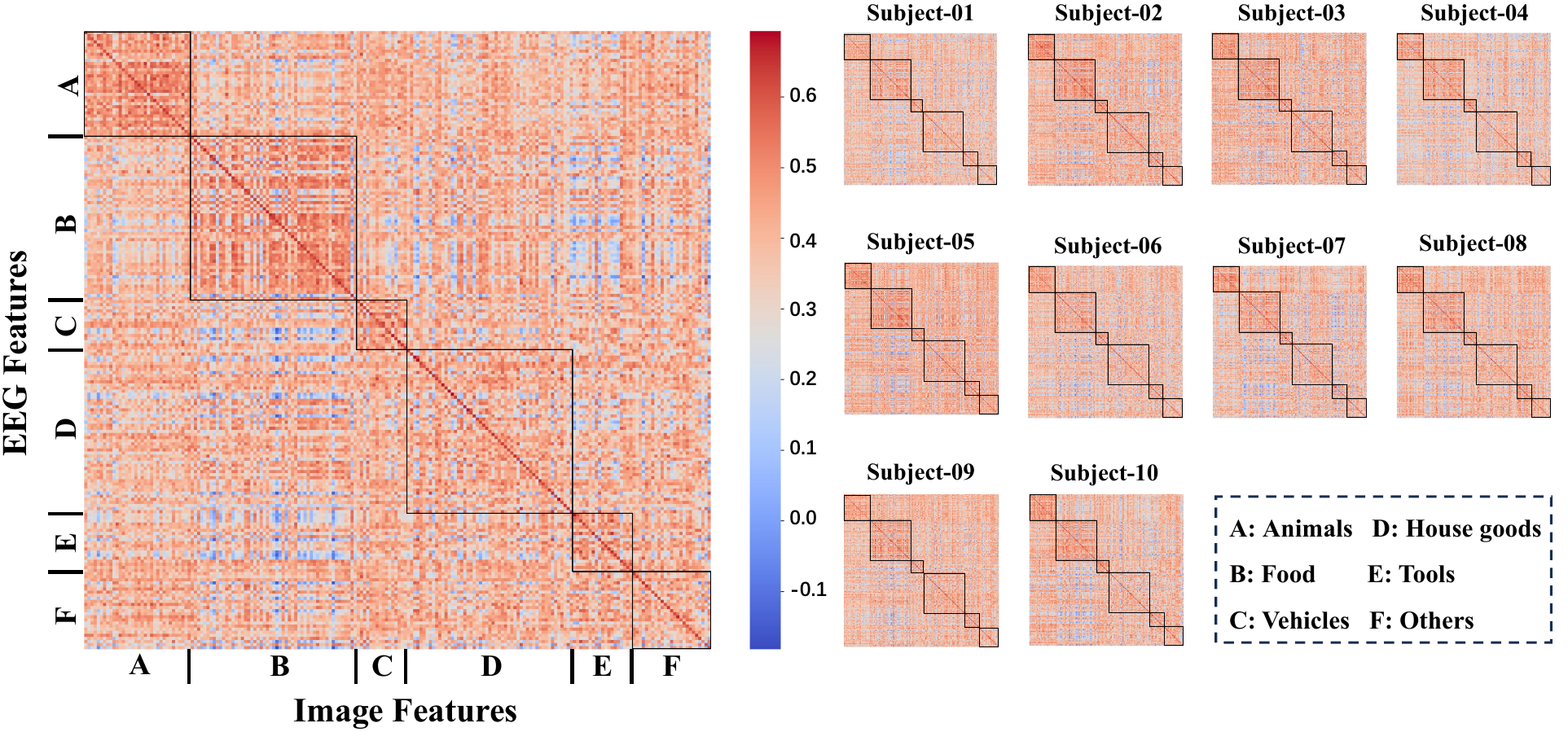}
    \caption{Cross-modal similarity matrices between EEG and image features on the THINGS-EEG2 dataset.}
    \label{fig:simheatmap}
\end{figure}

Figure~\ref{fig:simheatmap} shows the cross-modal similarity heatmap on the THINGS-EEG2 dataset, illustrating cosine similarities between 200 EEG features and their corresponding image features, with similarity computed via Equation~\ref{eq:similatiryscorediv} and ~\ref{eq:similatiryscoreall}. To reveal semantic patterns, test samples are grouped into six categories: Animals, Food, Vehicles, House goods, Tools, and Others, and arranged accordingly. A clear diagonal indicates successful one-to-one alignment between modalities, while block-wise high similarity suggests consistent semantic preservation. 

\begin{figure}[htbp]
    \centering
    \includegraphics[scale=0.99]{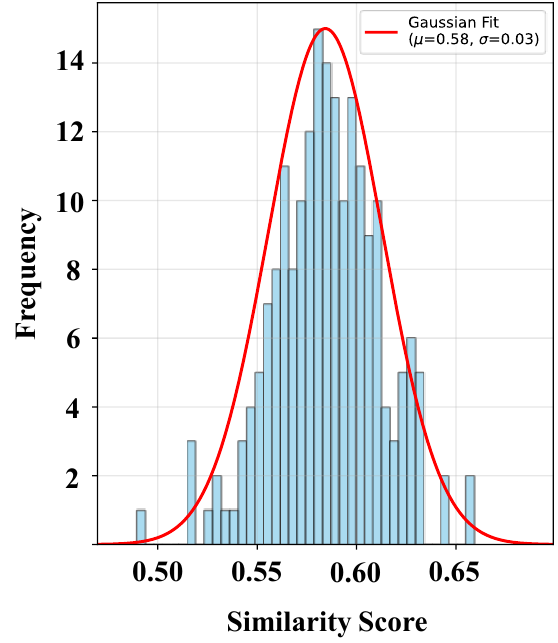}
    \caption{Distribution of ground truth similarity scores between EEG and image features.}
    \label{fig:simsta}
\end{figure}

Figure~\ref{fig:simsta} shows the distribution of ground truth similarity scores between EEG and image features. In subfigure (a), the histogram with a Gaussian fit ($\mu$ = 0.58, $\sigma$ = 0.03) indicates a near-normal distribution centered around a moderate-to-high similarity range. Subfigure (b) confirms this pattern with a compact boxplot and few outliers. These results suggest the ground truth similarities are consistent and form a reliable reference for evaluating alignment models. Notably, since we adopt a softened contrastive loss, the predicted similarity scores are less polarized compared to previous works, leading to a more continuous distribution between matched and unmatched pairs.

% Figure~\ref{fig:similarity_visualization} provides an overview of the predicted similarities between EEG and image features. As shown in subfigure (a), the similarity heatmap exhibits a strong diagonal structure, indicating accurate one-to-one alignment between paired EEG and image embeddings. Subfigure (b) shows the distribution of similarity scores, with most values concentrated between 0.50 and 0.56, suggesting consistent and well-calibrated cross-modal similarity estimation across samples.

% %visualizaiton_similarity
% \begin{figure}[htbp]
%     \centering
%     \includegraphics[width=0.9\textwidth]{simanalysisv1.pdf}
%     \caption{Visualizaiton of Similarity score}
%     \label{fig:similarity_visualization}
% \end{figure}

% \begin{figure}[htbp]
%     \centering
%     \includegraphics[width=\textwidth]{simditributev2.pdf}
%     \caption{Distribution of ground truth similarity scores between EEG and image features.}
%     \label{fig:simsta}
% \end{figure}

\subsection{Top-5 Retrieval Visualization}

\begin{figure}[htbp]
    \centering
    \includegraphics[width=\textwidth]{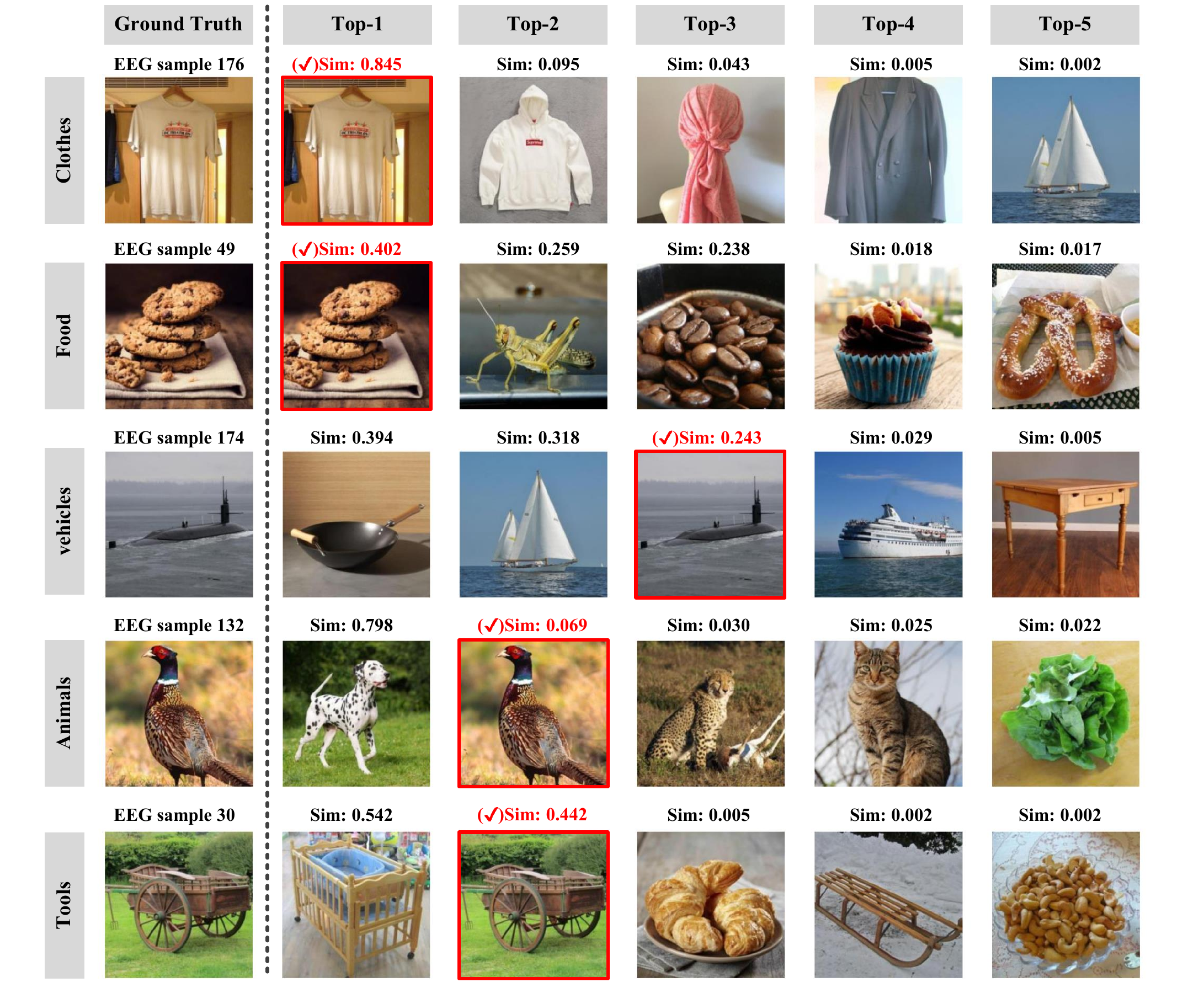}
    \caption{Top-5 retrieval visualization results for EEG samples.}
    \label{fig:top5_retrieval}
\end{figure}

Figure~\ref{fig:top5_retrieval} shows qualitative Top-5 retrieval examples. In most cases, the ground-truth image appears within the top ranks, indicating that the learned EEG-image representations are semantically aligned and retrieval-relevant.
%top-5visualization

\subsection{Discussion on Model Parameters and Computational Efficiency}
We further assess the parameter overhead and computational efficiency of our NeuroCLIP framework. As shown in Table~\ref{tab:param_comparison}, NeuroCLIP introduces only a marginal increase in total parameter count when built upon different CLIP-VIT backbones. Specifically, the added parameter percentage ranges from +1.57\% for VIT-B/16 to +0.68\% for VIT-H/14, demonstrating excellent scalability even when applied to large vision encoders.
In terms of runtime, we observe that the additional operations (e.g., dynamic filtering, prompt tuning) introduce a negligible cost. The difference in per-iteration execution time remains under 0.1 seconds, confirming that NeuroCLIP maintains high computational efficiency without compromising performance. Runtime statistics, including average time to process a single subject, time per training batch (with batch size 64), and retrieval time for evaluating 200 test samples, are also reported in Table~\ref{tab:param_comparison}. All measurements were conducted on a single NVIDIA A100 GPU.
%params numbers
\begin{table}[htbp]
\centering
\renewcommand{\arraystretch}{1.3}
\caption{Parameter Comparison of Different VIT Backbones with NeuroCLIP}
\label{tab:param_comparison}
\resizebox{\textwidth}{!}{
\begin{tabular}{lcccccc}
\toprule
\textbf{Vision Encoder} & \textbf{Params (M)} & \textbf{Emb Dim} & \textbf{NeuroCLIP (+\%)} & \textbf{Runtime on a single subject} & \textbf{Runtime per batch} & \textbf{Retrival time} \\
\midrule
VIT-B/16   & 86.19   & 512   & +1.57\%  & 6.53min  & 0.14s & 1.2s\\
VIT-B/32   & 87.85   & 512   & +1.54\%  & 7.22min  & 0.16s & 1.3s\\
VIT-L/14   & 303.97  & 768   & +0.86\%  & 78.82min & 1.84s &3.46s \\
VIT-H/14   & 632.08  & 1024  & +0.68\%  & 200.01min & 6.22s &5.99s\\
\bottomrule
\end{tabular}
}
\end{table}

\section{Conclusion and Future Work}

\subsection{Conclusion}

In this work, we introduced \textbf{NeuralCLIP}, a novel multimodal framework that bridges human neural activity and visual representations by adapting CLIP-style contrastive learning to the EEG domain. Unlike traditional approaches that treat EEG signals as a simple modality replacement, NeuralCLIP rethinks the prompt tuning paradigm through the lens of brain-inspired adaptation. Our dual-branch image encoder enables token-level fusion, allowing the model to dynamically reshape visual representations under neural constraints. Furthermore, we are the first to integrate visual prompt tokens into EEG–image alignment and demonstrate their effectiveness in a self-adaptive prompting mechanism. Coupled with a soft contrastive objective tailored to the semantic uncertainty of EEG signals, NeuralCLIP significantly improves zero-shot EEG-to-image retrieval on the THINGS-EEG2 dataset. These results highlight the viability of bringing foundation model principles to brain-computer interface research.

\subsection{Future Work}

While NeuralCLIP opens new directions for brain–vision alignment, several limitations and opportunities remain. First, our current EEG encoder is relatively shallow and task-specific. Future work could incorporate more expressive backbones, such as attention-based graph neural networks, to better capture spatiotemporal dependencies across electrodes.

Second, the self-prompting behavior in NeuralCLIP emerges implicitly through architectural design. A promising extension would be to make this behavior explicit by designing a learnable prompt controller—one that adjusts prompt tokens based on global EEG context, memory cues, or task semantics.

Third, although we use visual prompt tokens, the framework remains unimodal in its prompt representation. Inspired by large language models, future work could explore \textit{cross-modal prompting}, where learned EEG-derived prompts directly influence both vision and text encoders in a unified space, opening pathways toward brain-to-text generation or brain–image–language tri-alignment.

Lastly, as foundation models scale, aligning noisy, low-bandwidth signals like EEG to high-dimensional semantic spaces remains challenging. Our work suggests that incorporating physiological constraints into pretraining objectives, or using generative pretext tasks (e.g., masked prediction from EEG), may further enhance the generalization of neural-aligned multimodal systems.

\section{Acknowledgments}
This work was supported in part by the National Natural Science Foundation of China under Grant 62276169, in part by the Medical-Engineering Interdisciplinary Research Foundation of Shenzhen University under Grant 2024YG008, in part by the Shenzhen University-Lingnan University Joint Research Programme, in part by the Shenzhen-Hong Kong Institute of Brain Science-Shenzhen Fundamental Research Institutions under Grant 2023SHIBS0003, in part by the STI 2030-Major Projects 2021ZD0200500, and in part by the Open Research Fund of the State Key Laboratory of Brain-Machine Intelligence, Zhejiang University (Grant No. BMI2400008).

\bibliographystyle{unsrt}
\bibliography{reference.bib}

\end{document}